\documentclass[%
 preprint,
 amsmath,amssymb,
 aps,
]{revtex4-2}

\usepackage{mathtools}
\usepackage{graphicx}
\usepackage{bm}
\usepackage[mathlines]{lineno}
\usepackage{newtxtext}
\usepackage{newtxmath}
\usepackage{upgreek}
\usepackage{natbib}
\usepackage{hyperref}
\linenumbers\relax 

\begin{document}
\nolinenumbers

\preprint{Preprint}

\title{Field investigation of 3D snow settling dynamics under weak atmospheric turbulence}

\author{Jiaqi Li}
\author{Michele Guala}
\author{Jiarong Hong}%
 \email{jhong@umn.edu}

\affiliation{Saint Anthony Falls Laboratory, University of Minnesota, Minneapolis, MN, USA}
\affiliation{Department of Mechanical Engineering, University of Minnesota, Minneapolis, MN, USA}
\affiliation{Department of Civil, Environmental, and Geo- Engineering, University of Minnesota, Minneapolis, MN, USA}

\date{\today}

\begin{abstract}
Research on the settling dynamics of snow particles, considering their complex morphologies and real atmospheric conditions, remains scarce despite extensive simulations and laboratory studies. Our study bridges this gap through a comprehensive field investigation into the three-dimensional (3D) snow settling dynamics under weak atmospheric turbulence, enabled by a 3D particle tracking velocimetry (PTV) system to record over a million trajectories, coupled with a snow particle analyzer for simultaneous aerodynamic property characterization of four distinct snow types (aggregates, graupels, dendrites, needles). Our findings indicate that while the terminal velocity predicted by the aerodynamic model aligns well with the PTV-measured settling velocity for graupels, significant discrepancies arise for non-spherical particles, particularly dendrites, which exhibit higher drag coefficients than predicted. Qualitative observations of the 3D settling trajectories highlight pronounced meandering in aggregates and dendrites, in contrast to the subtler meandering observed in needles and graupels, attributable to their smaller frontal areas. This meandering in aggregates and dendrites occurs at lower frequencies compared to that of graupels. Further quantification of trajectory acceleration and curvature suggests that the meandering frequencies in aggregates and dendrites are smaller than that of morphology-induced vortex shedding of disks, likely due to their rotational inertia, and those of graupels align with the small-scale atmospheric turbulence. Moreover, our analysis of vertical acceleration along trajectories elucidates that the orientation changes in dendrites and aggregates enhance their settling velocity. Such insights into settling dynamics refine models of snow settling velocity under weak atmospheric turbulence, with broader implications for more accurately predicting ground snow accumulation. 
\end{abstract}

\maketitle


\section{\label{sec:intro}Introduction}

 Understanding the intricacies of snow settling dynamics is critical for accurately modeling snow accumulation, which has various scientific and socio-economic implications. These include issuing natural hazard warnings such as avalanches \citep{steinkogler2014influence} and snow-melt floods \citep{marks1998sensitivity}, understanding snow hydrology and its influence on local climates \citep{clark2011representing}, and optimizing traffic management during snow events \citep{ogura2002study}. A crucial determinant in the rate of snow accumulation is the settling velocity of snow particles, which can vary significantly, ranging from 0.5 m/s to speeds exceeding 3 m/s \citep{garrett2014observed,nemes2017snowflakes,li2021settling}. This variability greatly influences the drift distance of snowflakes as they descend from clouds to the ground. Presently, weather forecast models often struggle with precise predictions of ground snow accumulation, leading to potential economic repercussions \citep{winkler2015importance}. The variability in the settling velocity of snow particles in the atmosphere has been historically attributed to their morphology (e.g., size and shape), which poses a challenge in predicting their aerodynamic drag due to their complex and variable shapes \citep{locatelli1974fall,bohm1989general,tagliavini2021numerical,tagliavini2021drag}. Snow particle morphology is mainly determined by environmental conditions within clouds, such as temperature and humidity (i.e., supersaturation). The microphysics of ice crystal formation, extensively studied in works like \citet{magono1966meteorological} and \citet{libbrecht2005physics}, reveals a variety of emerging crystal shapes. These range from disk-like plates and dendrites to thin-cylinder needles and columns. In conditions of high supersaturation, small, supercooled droplets can adhere to these crystals through a process known as riming, leading to the creation of sphere-like graupels. As these ice crystals fall from clouds to the ground, inter-particle collisions occur, resulting in increasingly complex particle structures such as fragments and aggregates. Besides, the interaction between air turbulence and snow particle settling has been often overlooked in simulations and laboratory experiments. Atmospheric turbulence is typically sustained by the large velocity gradients of the high Reynolds number atmospheric surface layer, where coherent structures across various scales emerge, and modulate the snow settling velocity \citep[see][]{garrett2014observed,nemes2017snowflakes,li2021settling}.
 
Historically, measurements of snow particle fall speed did not account for the influence of atmospheric turbulence. The terminal fall speed, strictly defined in quiescent flow, was directly linked to aerodynamic drag and influenced by factors like particle size, shape, and mass. Various studies, including early research by \citet{nakaya1935simultaneous}, have sought to empirically correlate fall speeds with particle sizes. They observed an increase in velocity with size for graupels, crystals with droplets, and needles, while noting that dendrites and powder snow typically fall at a slower rate ($\sim$ 0.5 m/s), regardless of size. However, as their study was carried out in the laboratory setting, the snow particles might not reach their terminal velocity in a confined space. In a later study, \citet{heymsfield1972ice} introduced equations for calculating the terminal velocities of different snow morphologies, based on field measurements of drag coefficients, aspect ratios, and densities. Following this, \citet{locatelli1974fall} developed a specialized measurement instrument under a 3.8-meter-high shielded tower, to ensure snow particles reached terminal velocity during measurement. Their extensive collection of over 300 varied snow particles led to the development of empirical equations based on dimensional power laws, each tailored to specific snow morphologies and dependent on particle size. These studies underscore the importance of size and shape in determining the varying terminal velocities of snow particles. Despite these advancements, a comprehensive understanding of the detailed settling kinematics for these diverse morphologies remains an area for further exploration.

Kajikawa's extensive research from 1976 to 1997 laid a foundational understanding of snow particle dynamics, focusing on the free-falling behaviors of various snow particle types, such as columnar snow, early snow/aggregates, and plate-like snow \citep{kajikawa1976observation,kajikawa1982observation,kajikawa1989observation,kajikawa1992observations,kajikawa1997observations}. In these laboratory experiments, they documented a spectrum of free-fall motions, ranging from stable, horizontal movement-free descents to more complex patterns like non-rotating glides, swings, rotating glides, and spiral motions. Notably, the spiral motions exhibit inherent frequencies that correlate with the particle's Reynolds number, providing insights into the free-fall dynamics of snow particles. More systematic studies investigated the falling dynamics of idealized anisotropic particles, including disks and thin cylinders. These studies revealed that due to their large aspect ratios, such particles often orient themselves to maximize their projected area downwards during stable falls, i.e., preferential orientation. However, this steady fall is not always maintained; instabilities can lead to fluttering and even tumbling motions. These falling dynamics were explored extensively through experiments and simulations by researchers like \citet{willmarth1964steady}, \citet{auguste2013falling}, and \citet{tinklenberg2023thin}. Their work demonstrated the diverse falling styles of disks in quiescent flow, influenced by varying combinations of Reynolds number ($Re$) (or Galileo number, $Ga$; Archimedes number, $Ar$), and dimensionless moment of inertia ($I^*$). Similarly, thin cylinders, as studied by \citet{jayaweera1965behaviour} and \citet{toupoint2019kinematics}, exhibit comparable settling dynamics in quiescent flow. It was observed that due to their larger aspect ratio, even minor disturbances could induce more pronounced instabilities, leading to complex spinning (rotation around the axis of symmetry) and tumbling (rotation around other axes) in these particles. These movements are important as they affect the settling of these particles through the air, potentially changing their frontal area and their drag coefficient, which in turn influences their settling velocity. As a result, particle morphology and falling styles are deeply interconnected.

Air turbulence has been observed to modulate the settling velocity and spatial distribution of heavy inertial particles, regardless of their shape, simply due to their inability to follow exactly the motion of the fluid flow around them \citep{maxey1987gravitational,wang1993settling,yang1998role,aliseda2002effect,good2014settling,falkinhoff2020preferential}. Most studies have focused on point particles or small spherical particles, trying to separate morphological effects from turbulence effects. As anisotropic particles already exhibit various dynamics in quiescent flow, turbulence introduces more disturbances, suggesting that the two effects can hardly be decoupled \citep{voth2017anisotropic}. \citet{esteban2020disks} conducted experiments on free-falling disks and observed unique and complex settling behavior in turbulent flows (slow tumbling \& levitation). These motions displayed frequencies significantly lower than those of natural disks settling in still air. Interestingly, they noted an increase in settling velocity with greater horizontal velocity fluctuations and a decrease in oscillation frequency. Moreover, \citet{siewert2014orientation} conducted simulations on settling spheroids with various shape factors, including two extremes: disks (oblate spheroids) and needles (prolate spheroids), under various levels of turbulence. They observed that the preferential orientation of anisotropic particles is randomized by increasing level of turbulence, thus leading to more enhanced settling velocity (even though the morphology effect remained strong under weak turbulence). 

Despite the extensive numerical simulations and laboratory experiments, there remains a notable gap in field data that capture the complexity of realistic snow particles and atmospheric flow conditions, as compared to the usage of simplified model particles \citep{siewert2014orientation,toupoint2019kinematics,esteban2020disks,tinklenberg2023thin}, and controlled laboratory settings \citep{locatelli1974fall,kajikawa1982observation,kajikawa1989observation,kajikawa1992observations,kajikawa1997observations}. Therefore, field data is crucial for a deeper understanding of the settling dynamics of snow particles with varied morphologies in weakly turbulent conditions. Our group has been actively involved in field investigations of snow settling for the last decade. A significant advancement was the development of a super-large-scale particle image velocimetry system (SLPIV) by \citet{hong2014natural}. This system has been instrumental in visualizing flow structures in the wake of wind turbines \citep{hong2014natural,dasari2019near} and characterizing the atmospheric turbulent boundary layer \citep{toloui2014measurement,heisel2018spatial}. More recently, it has been applied to research on snow settling dynamics. \citet{nemes2017snowflakes} utilized this technology to quantify the settling trajectories of snow particles, measuring their Lagrangian velocity, acceleration, and aerodynamic properties. Their findings revealed a significant enhancement in settling velocity due to turbulence. Building on this, \citet{li2021settling} explored snow settling and clustering under various conditions, noting clustering at near-critical Stokes numbers and an increase in settling velocity correlating with concentration and cluster size. These findings indirectly support the preferential sweeping mechanism. Further, \citet{li2021evidence} provided direct evidence of preferential sweeping in atmospheric turbulence by simultaneously using SLPIV and PTV for flow and snow trajectory quantification. They observed increased snow concentration and enhanced settling velocity on the downward side of vortices, directly supporting the preferential sweeping mechanism. However, these studies were limited by planar imaging, which restricts the observation of snow particles' 3D motion, especially the spanwise motion, and did not consider the morphology effect of the snow particles. Therefore, comprehensive 3D field investigations and simultaneous, detailed measurements of snow morphology are essential.

In this study, we aim to bridge this gap by conducting field measurements during snow events using an imaging-based three-dimensional particle tracking velocimetry (3D PTV) system \citep{bristow2023imaging} for tracking 3D snow particle trajectories and a snow particle analyzer \citep{li2023snow} for assessing snow morphology and density. Our objectives are threefold: to understand how snow morphology influences snow aerodynamic properties, to determine the impact of morphology on particle 3D settling kinematics, and to assess how these dynamics affect snow settling velocity. Section \ref{sec:method} of this paper will detail the measurement instruments and data processing procedures. Section \ref{sec:result} will discuss the results and findings, followed by conclusions and discussions in Section \ref{sec:cd}.

\section{\label{sec:method}Method}

\begin{figure}
  \centerline{\includegraphics[scale = 0.9]{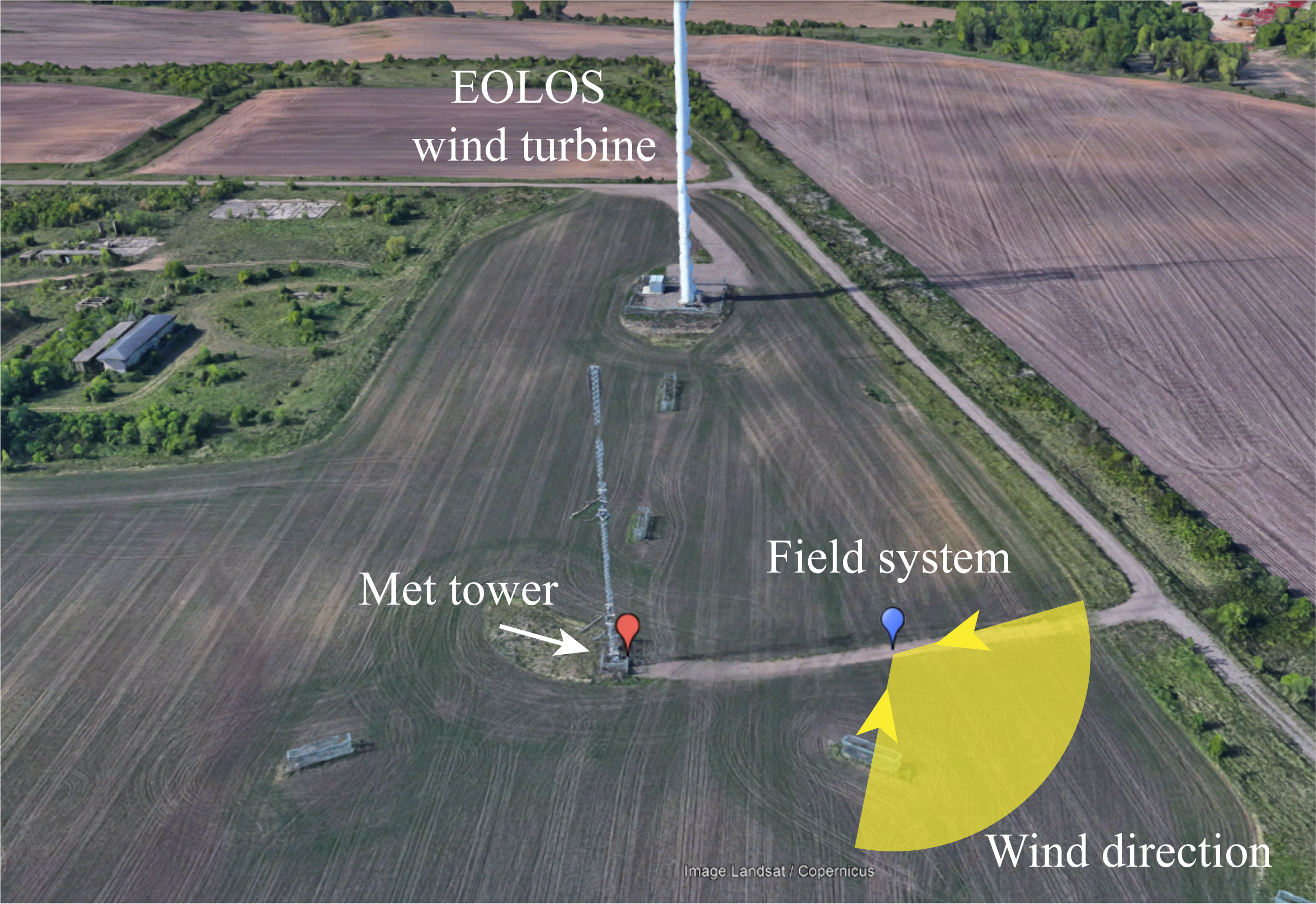}}
  \caption{Aerial view of the experimental field site in Rosemount, Minnesota, retrieved from Google Maps. The annotations in the satellite image highlight the location of the system deployment and the meteorological tower, with the transparent yellow circle segment indicating the range of wind directions during the field deployments.}
\label{fig:1}
\end{figure}

We conducted a series of field experiments at the EOLOS field research station (Figure \ref{fig:1}) in Rosemount, MN, USA, spanning the winter seasons from November 2021 to April 2023. The research station is well-equipped with a meteorological tower, which includes sensors for wind velocity, temperature, and humidity. These instruments are crucial for assessing the atmospheric and turbulent conditions during our field experiments. The tower is fitted with four sonic anemometers (CSAT3, Campbell Scientific) at heights of 10, 30, 80, and 129 m. These anemometers, with a 20 Hz sampling rate and path lengths of 5.8 cm horizontally and 10 cm vertically, provide detailed wind velocity data. Additionally, six cup-and-vane anemometers, each with a 1 Hz sampling rate, are positioned at elevations of 7, 27, 52, 77, 102, and 126 m to complement the wind measurements. In each field deployment, we utilized a three-dimensional particle tracking velocimetry (3D PTV) system, as described by \citet{bristow2023imaging}, to capture the trajectories of settling snow particles. To characterize the morphology and density of these snow particles, we employed a digital inline holography (DIH) system integrated with a high-precision scale, known as a snow particle analyzer, following the methodology outlined by \citet{li2023snow}.

\subsection{\label{sec:21}Experimental setup and data processing}

\begin{figure}
  \centerline{\includegraphics[scale = 0.9]{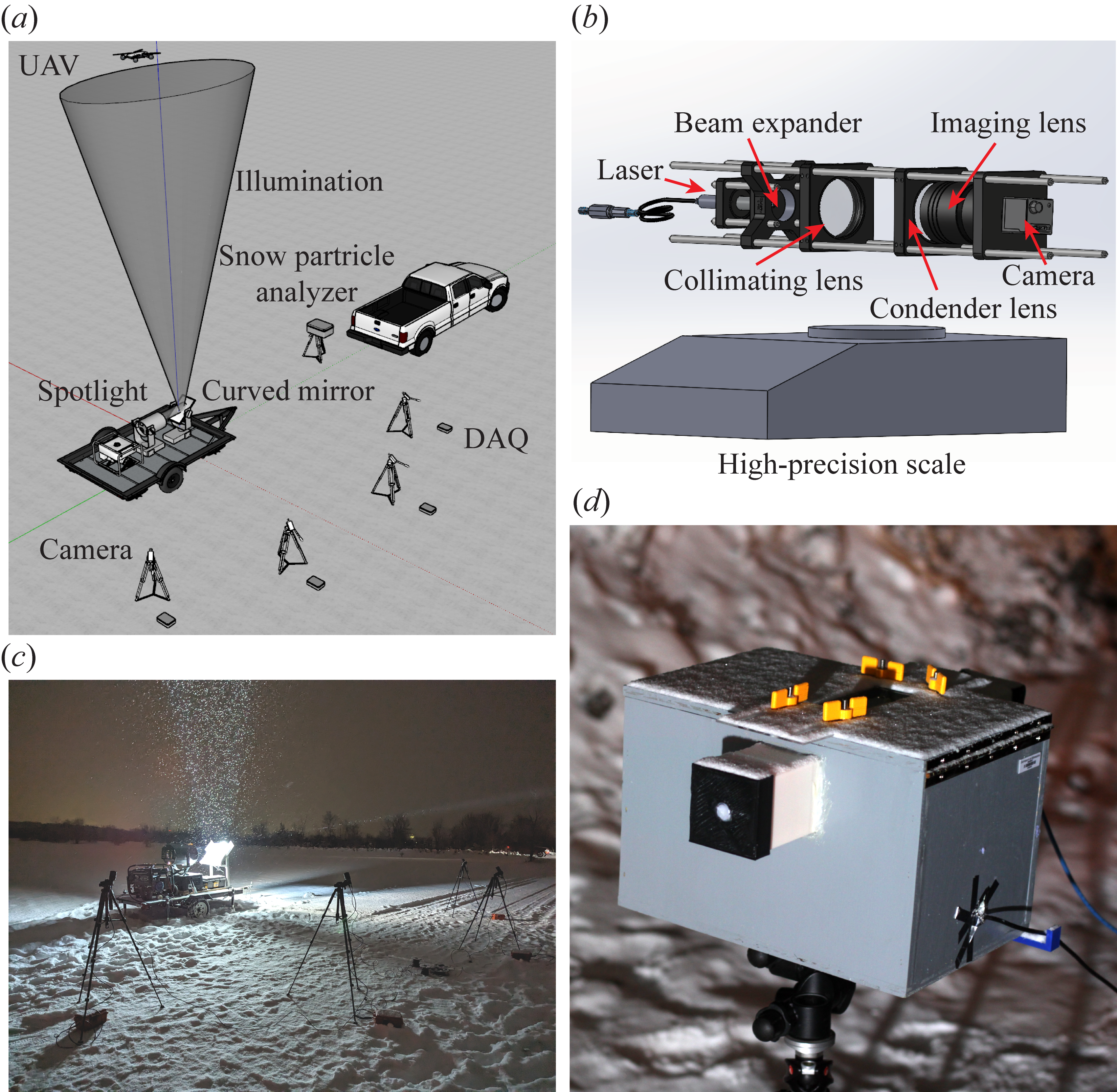}}
  \caption{(\textit{a}) Schematic depicting the field setup for the 3D particle tracking velocimetry (PTV) system, consisting of four cameras with their data acquisition units (DAQs), light source, and an unmanned aerial vehicle (UAV) for camera calibration, together with the snow particle analyzer. (\textit{b}) Design of the snow particle analyzer combining the digital inline holography system and a high-precision scale. Actual field deployment images show (\textit{c}) the 3D PTV system in operation at night and (\textit{d}) the snow particle analyzer for data collection.}
\label{fig:2}
\end{figure}

Figures \ref{fig:2}\textit{a} and \textit{c} illustrate the setup of our 3D PTV system. This system consists of four wire-synchronized cameras (Teledyne FLIR, FLIR Black Fly S U3-27S5C color unit with Sony IMX429 sensor: 1464 $\times$ 1936 pixels, 4.5 $\upmu\mathrm{m/px}$) strategically positioned around a light cone 5.5 m away, spanning a 90-degree angle range. This light cone is created by reflecting and expanding light from a searchlight using a curved mirror, similar to the setup used in our planar measurements. The cameras are tilted upward with 58-degree angles, leading them to image at a sample volume 10 m above ground. Each camera is then connected to its own data acquisition unit. These units are equipped with a board-level computer for issuing image capture commands to the cameras, a solid-state drive for storing both the system software and captured images, and a dedicated power supply. The cameras capture images with 2$\times$ decimation (732 $\times$ 968 pixels) to reach 200 Hz frame rate. As the standard checkerboard method cannot be applied to a field of view 10 m above ground after dark in the field, we use the wand calibration method described by \citet{theriault2014protocol} for camera calibration. We use two colored light-emitting diodes (LEDs), set at a fixed distance apart on a carbon fiber rod, to act as the ``wand" attached to an unmanned drone. This calibration process is conducted multiple times before and after each deployment to ensure the same field condition between camera calibration and snow particle imaging. We have developed custom-designed camera control software that synchronizes the image capturing process across all four cameras. The calibration of the cameras is conducted using the open-access software easyWand \citep{theriault2014protocol}, which involves capturing images of the two colored LEDs as they move within the imaging volume. The software utilizes the trajectories of the two LEDs from all four cameras to conduct the calibration, resulting in a final reprojection error within 0.25 pixels.

For tracking snow particle trajectories, we utilize an open-source implementation of the shake-the-box (STB) method \citep{tan2020introducing}. This version builds on the original STB method proposed by \citet{schanz2016shake}, with enhancements specifically in the identification and removal of ghost particles. These improvements make it particularly suitable for our field data, which feature relatively high noise levels and a large field of view. This approach enables our system to capture snow particle trajectories within a considerable volume of approximately $4 \times 4 \times 6$ $\mathrm{m}^3$. The system achieves a spatial resolution of 6.3 mm/voxel and a temporal resolution of 200 Hz, allowing for detailed and precise tracking of snow particle movements. The tracked snow particle trajectories are then re-oriented as a group to have the average streamwise direction as the $x$ direction. Thus, the $y$ direction is defined as the spanwise direction, and the $z$ direction is defined as the vertical direction. From these trajectories, we obtain the Lagrangian velocity, $\boldsymbol{u}=(u_x,u_y,u_z)$, the Lagrangian accelerations, $\boldsymbol{a}=(a_x,a_y,a_z)$, and the resulting curvature, $\kappa=\| \boldsymbol{u} \times \boldsymbol{a} \| / \|\boldsymbol{u}\|^3$, where $\times$ represents the cross product. We use the second-order central difference method to calculate the Lagrangian velocity (first-order derivative) and acceleration (second-order derivative). The approximation introduces inherent errors, $O\left(\Delta t^2\right)$, which depends directly on the time step and is relatively small. However, the positioning errors of the snow particles can propagate and magnify in the velocity and acceleration calculation. As discussed by \citet{schanz2016shake} and \citet{tan2020introducing}, the iterative particle reconstruction, shake-the-box tracking, and trajectory filtering techniques significantly refine and reduce positioning errors. We quantify the root mean square of the difference between trajectory positions before and after filtering to be 0.3 pixels. This reduction potentially compensates for the positioning errors inherited from camera calibration, resulting in smaller errors in velocity and acceleration calculations. Consequently, the actual uncertainties in measuring velocity and acceleration are primarily influenced by the selection of filter length, which ranges from $45 \pm 2$ frames (see Appendix \ref{appA}). This leads to an average acceleration uncertainty of 0.34 $\mathrm{m/s^2}$.

To complement our 3D PTV system, we also deployed a snow particle analyzer near the 3D PTV setup to assess the morphology and density of snow particles during each snow event (Figure \ref{fig:2}\textit{b} and \textit{d}). All these measurements are crucial for accurately estimating the terminal velocity of snow particles in still air. As shown in Figure \ref{fig:2}\textit{c}, the snow particle analyzer employs a digital inline holography (DIH) system, which captures holograms of snow particles within a sample volume of $2.9 \times 2.2 \times 14.0$ $\mathrm{cm}^3$. This system achieves a spatial resolution of 14 $\upmu$m/pixel and a temporal resolution of 50 Hz. Through image analysis of the holograms, we obtain detailed information on particle size and shape, specifically the area equivalent diameter ($D_\mathrm{eq}$), major axis length ($D_\mathrm{maj}$), minor axis length ($D_\mathrm{min}$), area ($A_e$), etc. We also classify the shape of each particle into one of six types: aggregates, graupels, dendrites, plates, needles, and small particles. We define the characteristic particle size ($D_p$) as the area equivalent diameter for aggregates, graupels, and small particles, and as the major axis length for dendrites, plates, and needles. Additionally, a high-precision scale measures the weight of snow particles passing through the DIH sample volume, allowing us to estimate the average density of the particles. We also perform conditional sampling to achieve measurement of the density of individual snow particles.

For estimating the aerodynamic properties of snow particles, we follow the method proposed by \citet{bohm1989general}. This method involves calculating the Best number $X$ (also known as the Davies number), a dimensionless number that incorporates only the physical properties of snow particles and ambient air, and represents the equilibrium between gravity and drag forces. The Best number is defined as:
\begin{equation}
X=C_D Re_{p}^2=\frac{8 \rho_p V_p g \rho_a}{\pi \mu^2}\left(\frac{A}{A_{\mathrm{e}}}\right)^{1 / 4}. 
\label{eqn:1}
\end{equation}
Noteworthily, unlike particle Reynolds number ($Re_p=\rho_a W_0 D_p / \mu$) and drag coefficient ($C_D$), the definition of Best number eliminates the need to incorporate particle terminal velocity in the formulation, which is not readily available for complex snow particles. In equation \ref{eqn:1}, $\rho_p$ and $V_p$ are the density and volume of the snow particles, respectively, which are specific to the type of snow particle, as detailed in \citet{li2023snow}. Specifically, we approximate the complex snow particles as spheroids (graupels and small particles), combinations of small spheroids (aggregates), disks with thickness in correlation with their diameter (plates and dendrites), and thin cylinders (needles and columns). Such a method minimizes errors in volume estimation as compared to the typical spherical assumption used in the snow measurement community, resulting in the uncertainties of the volume estimation within 10$\%$ for snow particles with irregular shape (aggregates) and uncertainties of density within 20$\%$ for all demonstration cases in \citet{li2023snow}.  $A_e$ is the effective snow particle imaged area, and $A$ is the circumscribed area of the enclosing circle or ellipse. Such an area ratio, $A / A_e$, serves as a simplified 2D measure of porosity and is instrumental in better predicting the drag of complex snow particles. As the snow morphological parameters are quantified by the snow particle analyzer while particles settle in various orientations, we assume that the ratio $A/A_e$ remains constant regardless of orientation. Finally, $\rho_a$ and $\mu$ are the density and viscosity of air, respectively.

Following the definition of the Best number, the drag coefficient of snow particles is modeled as a function of the particle Reynolds number, accounting for the unique morphology of snow particles. This approach indirectly incorporates the effect of snow particle density, which contributes to increasing the settling velocity. According to Stokes' law for $Re_p \sim O(1)$, the correlation for the drag coefficient dependent on the particle Reynolds number is $C_D = 24/Re_p$. However, the Stokes' law becomes invalid as the Reynolds number increases, especially for complex snow particles. Researchers have made various attempts to model the drag coefficient of snow particles theoretically \citep{bohm1989general,khvorostyanov2002terminal,khvorostyanov2005fall,mitchell2005refinements}. As suggested by \citet{bohm1989general} and references therein, the drag coefficient of snow particles is modeled by considering the boundary layer surrounding the snow particles as a whole:
\begin{equation}
C_D=C_0\left(1+\frac{\delta_0}{R e_p^{1 / 2}}\right)^2 ,
\label{eqn:2}
\end{equation}
where $C_0=0.6$ is an inviscid drag coefficient and $\delta_0=5.83$  is a parameter controlling the evolution of the particle boundary layer, likely depending on the particle surface roughness, both empirically estimated. Equation \ref{eqn:2} has the form of corrected Stokesian drag for a rigid sphere \citep{kaskas1970schwarmgeschwindigkeiten}, but with different coefficients, modulating the transition from a linear drag at low $Re_p$ to a constant drag coefficient $C_0$ in the $Re_p$ independent regime. The effect of different snow morphologies is included in the $A/A_e$ term in equation \ref{eqn:1}, which is then used to predict the snow type specific drag coefficients, $C_{De}=\left(A/A_e\right)^{3/4}C_D$. The $C_0$ and $\delta_0$ parameters have been more recently updated, along with the dependency on the area ratio, by \citet{heymsfield2010advances} and \citet{mccorquodale_trail_2021}.
Additional corrections considering turbulent boundary layer, temperature, humidity, and accounting for different snow particle types, have been discussed in \citet{khvorostyanov2002terminal,khvorostyanov2005fall} and \citet{mitchell2005refinements}.

As described in \citet{bohm1989general}, we then obtain a semi-analytical and semi-empirical equation for the particle Reynolds number by combining equations \ref{eqn:1} and \ref{eqn:2}:
\begin{equation}
R e_p=\frac{\delta_0^2}{4}\left(\left(1+\frac{4 X^{1 / 2}}{\delta_0^2 C_0^{1 / 2}}\right)^{1 / 2}-1\right)^2 .
\label{eqn:3}
\end{equation}
The terminal velocity of the snow particles ($W_0$) in quiescent air is then calculated from the Reynolds number, $R e_p=\rho_a W_0 D_p / \mu$. Once the terminal velocity is obtained, the aerodynamic particle response time is defined as $\tau_p=W_0/g$.

The analyses described have been meticulously applied to each snow particle type, leveraging the unique physical properties of individual particles, captured by the snow particle analyzer. Through a detailed examination of the collected holograms, we identify and classify each particle, subsequently analyzing their specific inertial properties, namely $D_p$, $\rho_p$, $A$, and $A_e$. By employing these properties within the Böhm model \citep{bohm1989general}, we were able to estimate the aerodynamic properties of each particle. This rigorous method allows us to calculate the distribution and mean values of the terminal velocity ($W_0$) and drag coefficient for individual snow particles and specific snow types.

\subsection{\label{sec:22}Turbulence and snow conditions in the field}

Over the course of the winter seasons from December 2021 to April 2023, we successfully carried out eight field deployments, encompassing a diverse range of environmental conditions. These deployments allowed us to study four major types of snow particles: aggregates, graupels, dendrites/plates, and needles/columns. We encountered wind speeds varying from a gentle 0.6 m/s to a more intense 8.4 m/s. Based on these wind speeds, we categorized the conditions into three turbulence levels: weak turbulence (wind speed less than 3 m/s with turbulent kinetic energy, TKE, below 0.3 $\mathrm{m}^2/\mathrm{s}^2$), moderate turbulence (wind speed between 3 and 6 m/s with TKE ranging from 0.3 to 2.0 $\mathrm{m}^2/\mathrm{s}^2$), and relatively strong turbulence (wind speed exceeding 6 m/s with TKE above 2.0 $\mathrm{m}^2/\mathrm{s}^2$). These turbulent properties were measured using the sonic anemometer positioned at a height of 10 m. Details of the estimation methods of these quantities can be found in \citet{li2021evidence}. We use the second-order structure function of the streamwise velocity fluctuation to estimate the dissipation rate ($\varepsilon$). The Taylor microscale ($\lambda$) is then calculated as $\lambda=u^{\prime} \sqrt{15 \nu / \varepsilon}$, where $u^{\prime}=\sqrt{\left(u_x^{\prime 2}+u_y^{\prime 2}+u_z^{\prime 2}\right) / 3}$ is the representative scale of fluctuating velocity, and $\nu$ is the viscosity of air. Given the variety of snow particle types and wind speeds, our field data encompasses a total of 31 distinct conditions. To effectively separate the influences of snow morphology and atmospheric turbulence on snow settling velocity, a more systematic classification of the field snow and turbulence conditions is essential. We propose using the settling parameter $S v_L=W_0 / u^{\prime}$, which quantifies the relative impact of turbulence on snow gravitational settling \citep{petersen2019experimental,brandt2022particle}. This parameter represents the ratio of the snow particle’s terminal velocity in still air ($W_0$) to the root-mean-square of the turbulent velocity fluctuations ($u^\prime$). A higher value of $Sv_L$ indicates that the influence of turbulence on the snow settling velocity is relatively minor.

\begin{table}\centering
\begin{tabular}{|l|c|c|c|c|c|c|c|c|c|c|c|c|c|c|}
\hline Dataset & \begin{tabular}{c}
$U$ \\
$\mathrm{~m} / \mathrm{s}$
\end{tabular} & \begin{tabular}{c}
$u_{x, \mathrm{rms}}$ \\
$\mathrm{m} / \mathrm{s}$
\end{tabular} & \begin{tabular}{c}
$u_{y, \mathrm{rms}}$ \\
$\mathrm{m} / \mathrm{s}$
\end{tabular} & \begin{tabular}{c}
$u_{z, \mathrm{rms}}$ \\
$\mathrm{m} / \mathrm{s}$
\end{tabular} & \begin{tabular}{c} 
TKE \\
$\mathrm{m}^2 / \mathrm{s}^2$
\end{tabular} & \begin{tabular}{c}
$L$ \\
$\mathrm{~m}$
\end{tabular} & \begin{tabular}{c}
$\tau_L$ \\
$\mathrm{~s}$
\end{tabular} & \begin{tabular}{c}
$\varepsilon$ \\
$\mathrm{cm}^2 / \mathrm{s}^3$
\end{tabular} & \begin{tabular}{c}
$\eta$ \\
$\mathrm{mm}$
\end{tabular} & \begin{tabular}{c}
$\tau_\eta$ \\
$\mathrm{s}$
\end{tabular} & \begin{tabular}{c}
$\lambda$ \\
$\mathrm{mm}$
\end{tabular} & \begin{tabular}{c}
$R e_\lambda$ \\
-
\end{tabular} & \begin{tabular}{c}
$T$ \\
${ }^{\circ} \mathrm{C}$
\end{tabular} & \begin{tabular}{c}
$\mathrm{RH}$ \\
$\%$
\end{tabular} \\
\hline Aggregate & 1.30 & 0.19 & 0.17 & 0.11 & 0.05 & 2.5 & 13 & 2.3 & 1.8 & 0.24 & 179 & 950 & -4 & 95 \\
\hline Graupel & 1.26 & 0.20 & 0.19 & 0.16 & 0.06 & 3.1 & 15 & 2.8 & 1.7 & 0.22 & 172 & 1114 & -10 & 87 \\
\hline Dendrite & 1.69 & 0.30 & 0.27 & 0.19 & 0.14 & 2.4 & 8 & 9.2 & 1.3 & 0.12 & 138 & 1139 & 0.4 & 96 \\
\hline Needle & 1.30 & 0.19 & 0.17 & 0.11 & 0.05 & 2.5 & 13 & 2.3 & 1.8 & 0.24 & 179 & 908 & -4 & 95 \\
\hline
\end{tabular}
  \caption{Overview of atmospheric turbulence parameters across datasets, detailing wind speed, turbulence fluctuations, turbulence kinetic energy (TKE), integral scale ($L$), dissipation rate ($\epsilon$), Kolmogorov scale ($\eta$), Taylor microscale ($\lambda$), Reynolds number ($Re_\lambda=\lambda u_{x,\mathrm{rms}} / \nu$), ambient temperature ($T$), and relative humidity (RH).}
  \label{tab:1}
\end{table}

\begin{figure}
  \centerline{\includegraphics[scale = 0.9]{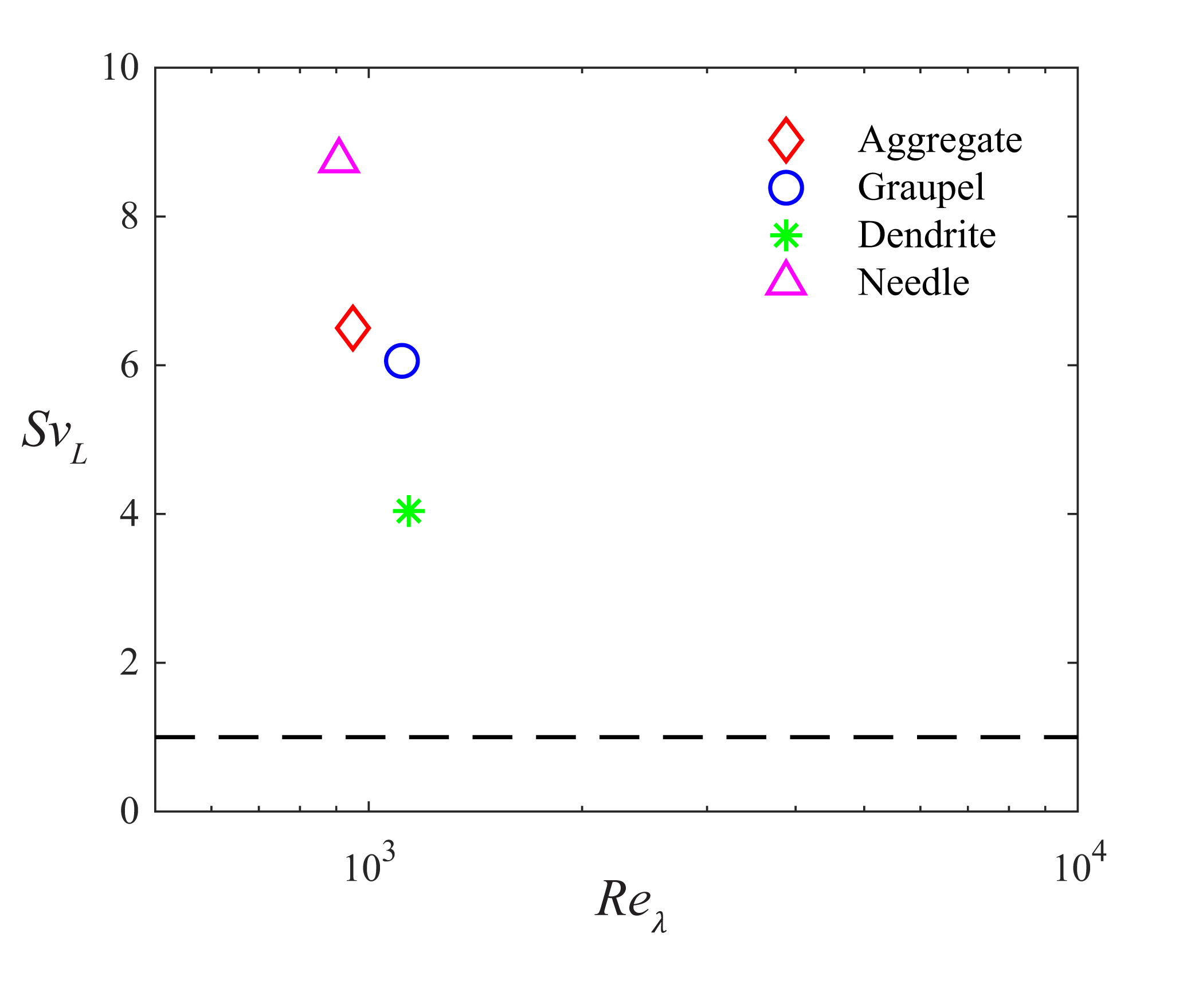}}
  \caption{A summary of the Taylor Reynolds number of the atmospheric flow ($Re_\lambda$) and settling parameter of the snow particles ($Sv_L$) for different snow particle types. Data points for aggregates are marked with red diamond, graupel with blue circle, dendrites with green star, and needles with magenta triangle.}
\label{fig:3}
\end{figure}

\begin{figure}
  \centerline{\includegraphics[scale = 0.85]{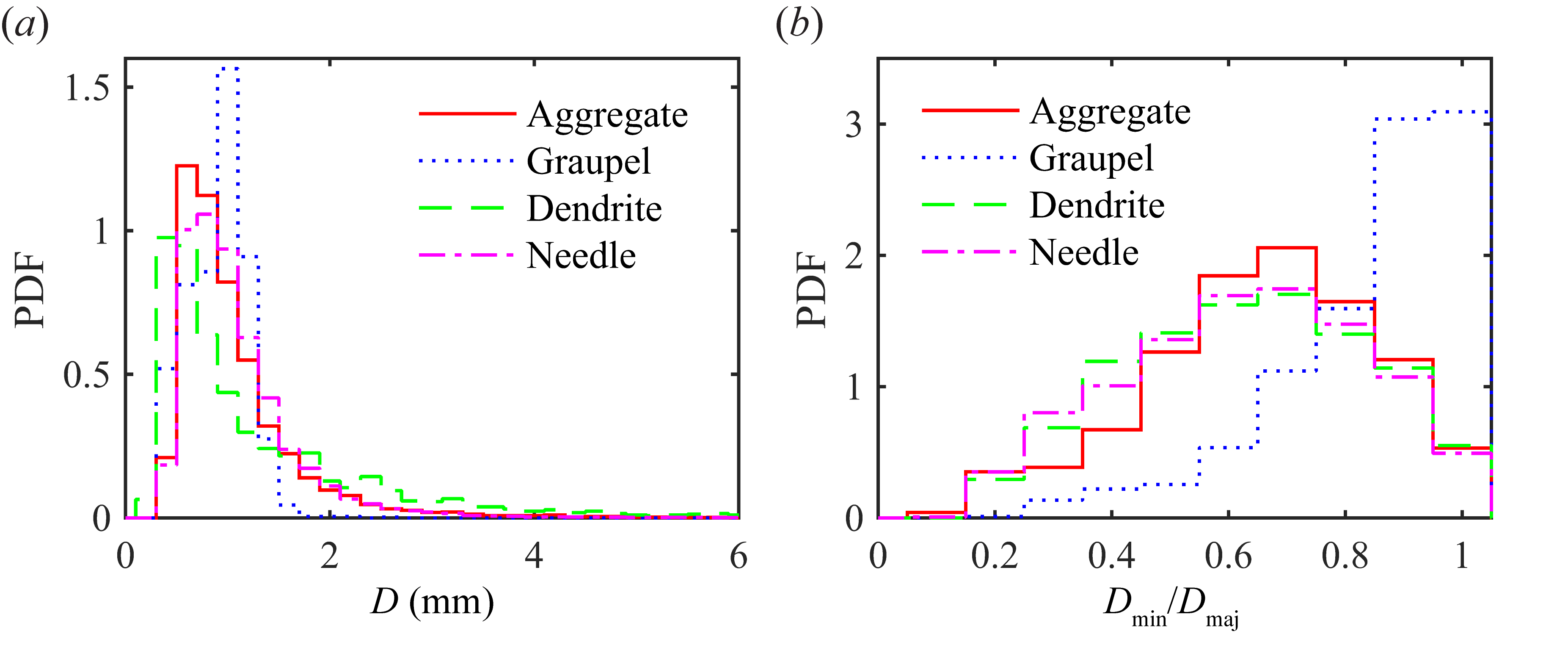}}
  \caption{(\textit{a}) Probability distribution functions (PDFs) of snow particle size ($D_p$, defined as the equivalent diameter for aggregates and graupels, as well as the major axis length for dendrites and needles) and (\textit{b}) PDFs of the aspect ratio ($D_\mathrm{min}/D_\mathrm{maj}$)  for various snow types, plotted with different line styles and colors: aggregates are represented by red solid lines, graupel by blue dotted lines, dendrites by green dashed lines, and needles by magenta dash-dotted lines.}
\label{fig:4}
\end{figure}

In our analysis, we utilized the settling parameter, wind speed, and turbulent kinetic energy (TKE) as key criteria to categorize our 3D PTV and snow particle analyzer datasets. This approach led us to identify four distinct groups, which we labeled as ‘weak turbulence’ cases, with relatively smaller Taylor Reynolds number ($Re_\lambda$) and higher settling parameters ($Sv_L$) as shown in Figure \ref{fig:3}. We assess the turbulence and micro-meteorological conditions for each group detailed in Table \ref{tab:1}, employing estimation methods as outlined in the studies by \citet{nemes2017snowflakes} and \citet{li2021evidence}. Each group is dominated by one specific type of snow particle, which constitutes more than half of the snow population in the dataset. These types are aggregates, graupels, dendrites, and needles, as detailed in Table \ref{tab:2} and illustrated in the size and shape distributions in Figure \ref{fig:4}. We leverage the capabilities of the snow particle analyzer, as detailed in Section \ref{sec:21}, to estimate these physical properties of snow particles. During a selected one-hour period characterized by dominant snow particle types, our analysis encompasses 200,000 holograms for each type of snow particle. This comprehensive dataset yields detailed information about approximately 28,000 aggregates, 13,000 graupels, 30,000 dendrites, and 21,000 needles. Complementing the snow particle analyzer measurements, our 3D PTV datasets include a total of 500 seconds of images for each dominant snow type, which are broken down into 50-second segments throughout the one-hour period selected. This rich dataset facilitates the identification of millions of snow particle trajectories, specifically around 322,000 for aggregates, 285,000 for graupels, 1,037,000 for dendrites, and 182,000 for needles, providing orders of magnitude more data than our previous studies. Specifically, our 3D PTV system measures the complete 3D velocity and particle acceleration components. The additional spanwise dimension of the data, compared to planar measurements, enables a thorough analysis of snow particle kinematics, including trajectory curvature and meandering. Furthermore, the integration of the 3D PTV system with the snow particle analyzer allows us to correlate the specific morphology of snow particles (e.g., size, shape, and type) with their settling behavior.

For the detected snow particles, their size and shape are measured through image analysis described in Section \ref{sec:21}. Given that these particles may present various orientations relative to the imaging plane, relying solely on the projected area (or equivalent diameter) falls short of providing a precise representation of the characteristic size of each particle, especially the non-spherical ones. We thus define the particle size as the equivalent diameter for aggregates and graupels, the major axis for dendrites (diameter) and needles (length), as detailed in \citet{li2023snow}.  Upon closer examination, we observed notable differences among these types. Graupels and needles, for instance, tend to have a more uniform size distribution, with a smaller average size and standard deviation compared to aggregates and dendrites. Aggregates and dendrites, on the other hand, are generally larger, and their datasets include a mix of other particle types, resulting in a broader size distribution. We also analyzed the aspect ratio (i.e., the ratio between the minor and major axes lengths, $D_\mathrm{min}/D_\mathrm{maj}$) of these snow particles, defined as the ratio of their minor to major axis lengths, as measured by the snow particle analyzer. Graupels predominantly exhibit aspect ratios greater than 0.8, indicating their near-spherical shape. In contrast, the aspect ratios for the other types vary significantly from one, suggesting more anisotropic shapes. In this respect, note that the 2D holograms do not allow to accurately capture the averaged thickness of plate-like crystals due to the random particle orientation, unless further analysis is performed on selected particle images as in \citet{li2023snow}. Furthermore, we measured the average density ($\overline{\rho_p}$), together with the average particle size ($\overline{D_p}$) and aspect ratio ($\overline{D_\mathrm{min}/D_\mathrm{maj}}$), of the four datasets using the snow particle analyzer. Needles, being solid crystals with minimal riming, have the highest average density of 360 $\mathrm{kg}/\mathrm{m}^3$. Dendrites follow with an average density of 280 $\mathrm{kg}/\mathrm{m}^3$, as it is influenced by the gaps between branches, which contribute to the overall porosity of the particles. Graupels have an average density of around 220 $\mathrm{kg}/\mathrm{m}^3$, aligning with our previous measurements. Moreover, aggregates exhibit the lowest density of around 90 $\mathrm{kg}/\mathrm{m}^3$, as expected, attributable to their larger size and higher porosity. 

\begin{table}
  \begin{center}
\def~{\hphantom{0}}
    \begin{tabular}{|c|c|c|c|c|c|c|c|c|}
\hline    Snow type & \begin{tabular}{c} Aggregate \\ $\%$ \end{tabular} & \begin{tabular}{c} Graupel \\ $\%$ \end{tabular} & \begin{tabular}{c} Dendrite \\ $\%$ \end{tabular} & \begin{tabular}{c} Needle \\ $\%$ \end{tabular} & \begin{tabular}{c} $\overline{D_p}$ \\ $\mathrm{~mm}$ \end{tabular} & \begin{tabular}{c} $\overline{D_{\text {min}} / D_{\text {maj}}}$ \\ - \end{tabular} & \begin{tabular}{c} $\overline{A/A_e}$ \\ - \end{tabular} & \begin{tabular}{c} $\overline{\rho_p}$ \\ $\mathrm{kg} / \mathrm{m}^3$ \end{tabular} \\ 
\hline    Aggregate & 55.8 & 21.1 & 14.0 & 9.1 & $0.96 \pm 0.63$ & $0.61 \pm 0.19$ & 1.22 & 90 \\
\hline    Graupel & 2.5 & 93.3 & 3.7 & 0.5 & $0.75 \pm 0.28$ & $0.80 \pm 0.16$ & 1.06 & 220 \\
\hline    Dendrite & 16.9 & 19.3 & 55.4 & 8.4 & $0.93 \pm 0.66$ & $0.58 \pm 0.21$ & 1.98 & 280 \\
\hline    Needle & 13.2 & 22.4 & 7.2 & 57.2 & $0.74 \pm 0.34$ & $0.58 \pm 0.20$ & 1.21 & 360 \\ \hline
    \end{tabular}
  \caption{Comparative overview of snow particle characteristics across various dataset groups, including the proportion of snow types where the dominant type exceeds 50$\%$ occurrence, the mean diameter (defined as the average equivalent diameter for aggregates and graupels, as well as the average major axis length for dendrites and needles), aspect ratio, area ratio, and density values characterizing each dataset group.}
  \label{tab:2}
  \end{center}
\end{table}

\section{\label{sec:result}Results}

Utilizing the snow particle analyzer, we have successfully measured both the morphology and density of snow particles, enabling us to accurately predict their aerodynamic properties. Additionally, our 3D PTV system has provided detailed 3D settling dynamics from millions of snow particle trajectories. Armed with this comprehensive data, we address three key questions in the following section: First, how does the morphology of snow particles influence their aerodynamic properties? Second, in what ways does morphology impact the settling kinematics of these particles? And third, how do the varying settling dynamics among different types of snow particles affect the overall settling velocity of snow? These inquiries form the core of our investigation, shedding light on the intricate interplay between snow particle morphology and their settling behavior through the atmosphere. 

\subsection{\label{sec:31}Aerodynamic properties}

This section presents an in-depth examination of the aerodynamic characteristics, including their terminal velocity, drag coefficient, and settling velocity, for each snow particle type. Table \ref{tab:3} consolidates key aerodynamic parameters derived from our analysis: the average settling velocity ($\overline{W_s}$) obtained through 3D particle tracking velocimetry (3D PTV), the average estimated still-air terminal velocity ($\overline{W_0}$) as outlined in Section \ref{sec:21}, the velocity fluctuation ($u^\prime$) and the Kolmogorov time scale ($\tau_\eta$) of the flow, the particle's Stokes number ($St_\eta=\tau_p/\tau_\eta$), their settling parameter ($Sv_L=\overline{W_0}/u^\prime$), and the Froude number ($Fr_\eta=a_\eta/g$, where $a_\eta=u_\eta/\tau_\eta$ is the Kolmogorov scale acceleration). Needles exhibit the highest terminal velocity among all four types. With the same particle size, the cylindrical-shaped needles have the smallest projected area and the highest density, leading to larger terminal velocities. The Stokes number gauges the particle's velocity response to sudden changes in flow, with values around one signifying a critical condition for turbulence-particle interactions. Settling parameters greater than one imply a weak influence of turbulence on the settling particles. The Froude number, a ratio of the characteristic flow acceleration ($a_\eta=u_\eta/\tau_\eta$) to gravitational acceleration, suggests that gravitational settling is more pronounced than the turbulence effect on the particles \citep{bec2014gravity}. Comparatively, the settling velocity enhancements from the terminal velocities are moderate, ranging up to 32$\%$ for aggregates, 13$\%$ for dendrites, 4$\%$ for needles, and 3$\%$ for graupels. These findings indicate that the turbulence effects (e.g., preferential sweeping and loitering) on particle settling is generally weak under the examined conditions. Variations in settling enhancement across snow types may be largely attributable to differences in particle size, shape, and density.

\begin{table}
  \begin{center}
\def~{\hphantom{0}}
    \begin{tabular}{|c|c|c|c|c|c|c|c|}
\hline    \begin{tabular}{c} Snow type \end{tabular} & \begin{tabular}{c} $\overline{W_s}$ \\ m/s \end{tabular} & \begin{tabular}{c} $\overline{W_0}$ \\ m/s \end{tabular} & \begin{tabular}{c} $u^{\prime}$ \\ m/s \end{tabular} & \begin{tabular}{c} $\tau_\eta$ \\ s \end{tabular} & \begin{tabular}{c} $S t_\eta$ \\ - \end{tabular} & \begin{tabular}{c} $S v_L$ \\ - \end{tabular} & \begin{tabular}{c} $F r_\eta$ \\ - \end{tabular} \\
\hline    aggregate & 1.37 $\pm$ 0.22 & 1.04 $\pm$ 0.31 & 0.16 & 0.238 & 0.44 & 6.5 & 0.003 \\
\hline    Graupel & 1.06 $\pm$ 0.25 & 1.03 $\pm$ 0.34 & 0.17 & 0.218 & 0.48 & 6.0 & 0.006 \\
\hline    Dendrite & 1.14 $\pm$ 0.24 & 1.01 $\pm$ 0.60 & 0.25 & 0.119 & 0.87 & 4.0 & 0.009 \\
\hline    Needle & 1.45 $\pm$ 0.17 & 1.40 $\pm$ 0.66 & 0.16 & 0.238 & 0.60 & 8.7 & 0.003 \\ \hline
    \end{tabular}
  \caption{Summary of characteristic parameters for snow particles and atmospheric flow, encompassing average terminal ($\overline{W_0}$) and settling velocities ($\overline{W_s}$), Stokes number ($S t_\eta$), settling parameter ($S v_L$), flow velocity scale ($u^{\prime}$), Kolmogorov time scale ($\tau_\eta$), and Froude number ($F r_\eta$).}
  \label{tab:3}
  \end{center}
\end{table}

\begin{figure}
  \centerline{\includegraphics[scale = 0.85]{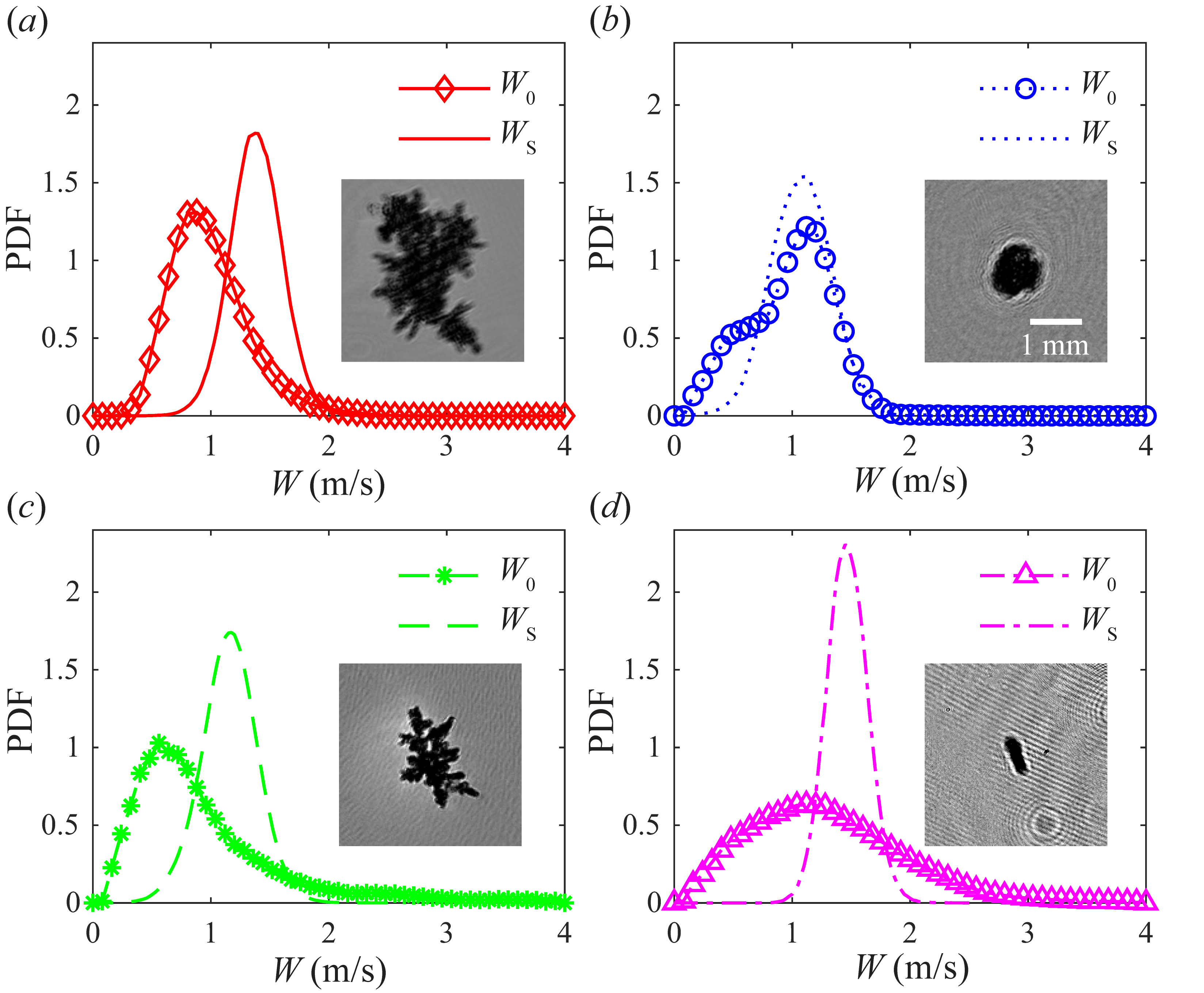}}
  \caption{Probability distribution functions (PDFs) contrasting the estimated still-air terminal velocity ($W_0$) using snow properties measured by the snow particle analyzer and the experimentally measured settling velocity ($W_s$) by the 3D PTV system for four datasets with different dominant snow types: (\textit{a}) aggregates, (\textit{b}) graupel, (\textit{c}) dendrites, and (\textit{d}) needles. Insets within each panel display representative holographic images of the corresponding snow particle type.}
\label{fig:5}
\end{figure}

Figure \ref{fig:5} presents a comparative analysis of the probability density functions (PDFs) for settling velocity ($W_s$) and estimated still-air terminal velocity ($W_0$) across various snow particle types. The estimate of $W_0$ is based on the Best number, $X=C_D Re_p^2$, which does not directly depend on the settling velocity of the snow particles, but rather on their physical properties and the ambient air. Following \citet{bohm1989general} approach, summarized by equations \ref{eqn:1}-\ref{eqn:3}, we estimate the terminal velocity from measurable geometric and inertial properties by the snow particle analyzer. The PDFs for graupel, which are nearly spherical in shape, exhibit a close overlap between the settling and terminal velocities, indicating a minimal influence of turbulent eddies on their settling dynamics. Note that the `dent' in the distribution of the terminal velocity of graupels in Figure \ref{fig:5}\textit{b} reflects their size distribution. On the contrary, for the other snow types—characterized by non-spherical geometries—the PDFs diverge despite the mean settling and terminal velocities for needles displaying only a 4$\%$ discrepancy. This variation suggests that the aerodynamic behavior of non-spherical particles is considerably affected by the randomization of their orientation due to flow disturbances and unsteady behavior. In quiescent conditions, particles falling stably tend to orient themselves to maximize the aerodynamic drag (i.e., preferential orientation), potentially due to the inertial forces of the surrounding media, presenting their maximal cross-sectional area perpendicular to the fall direction  \citep{willmarth1964steady,cho1981orientation}. However, in turbulent conditions, such a preferential orientation is not appreciable \citep{cho1981orientation,klett1995orientation}, and the varying orientations result in a reduced effective cross-sectional area, potentially leading to an increased average settling velocity for non-spherical particles. Furthermore, while the settling velocity distributions for different snow types approximate a Gaussian profile, the estimated terminal velocities are rather skewed. This asymmetry arises from the inherent size distributions of the snow particles, which are typically modeled using a gamma distribution \citep{field2007snow}.  We also acknowledge the potential sampling differences between the 3D PTV measurements (likely under-representing the finest size fraction) and the snow particle analyzer data collection.

Historical studies have demonstrated that the terminal velocity of snow particles exhibits a size-dependent characteristic, since the early research by \citet{nakaya1935simultaneous}, \citet{heymsfield1972ice}, and \citet{locatelli1974fall} fitting empirical data to establish a particle-mass-based approach to the settling. Specifically, \citet{locatelli1974fall} conducted a thorough investigation of various snow particle types, deriving power-law empirical formulas to represent the size-dependent terminal velocity, expressed as $W_0=a {D_p} ^b$, where $a$ and $b$ are constants that differ based on the snow particle type, based on shape and density. For our analysis, we employed the formulas relevant to aggregates of unrimed radiating assemblages of dendrite ($W_0=0.8D_p^{0.16}$), conical graupel ($W_0=1.2D_p^{0.65}$), rimed dendrites ($W_0=0.62D_p^{0.33}$), and rimed columns ($W_0=1.1L^{0.56}$, where $L$ is the length). Such power-law equations can be empirically obtained by fitting the size distributions in Figure \ref{fig:4} with the settling velocity in Figure \ref{fig:5}. We optimize the linear coefficient with the same exponent to impose the same mean and similar distribution of the settling velocity for each snow type, and thus obtained the empirical equations: $W_s=1.45D_p^{0.16}$ for aggregates, $W_s=1.2D_p^{0.65}$ for graupels, $W_s=0.92D_p^{0.33}$ for dendrites, $W_s=1.66L^{0.56}$ for needles. We thus obtain the same equation for $W_0$ and $W_s$ for graupels, suggesting close alignment in the mean values and distributions between the measured settling velocities and the estimated terminal velocities, the same as predicted using equations from \citet{bohm1989general}. This also confirms negligible effects by the specific atmospheric turbulence conditions monitored during the settling of graupels. In contrast, for other non-spherical types of snow, the linear coefficients for $W_s$ are higher than those of $W_0$. This discrepancy highlights the morphology effects that modulate the settling velocity of these non-spherical snow particles, with potential turbulence effects considering the varying particle orientation because of turbulence disturbances, and the production of the Stokes number and settling parameter reaching critical condition \citep[$St_\eta Sv_L \sim 1$, as suggested in][]{petersen2019experimental,brandt2022particle}.

\begin{figure}
  \centerline{\includegraphics[scale = 0.85]{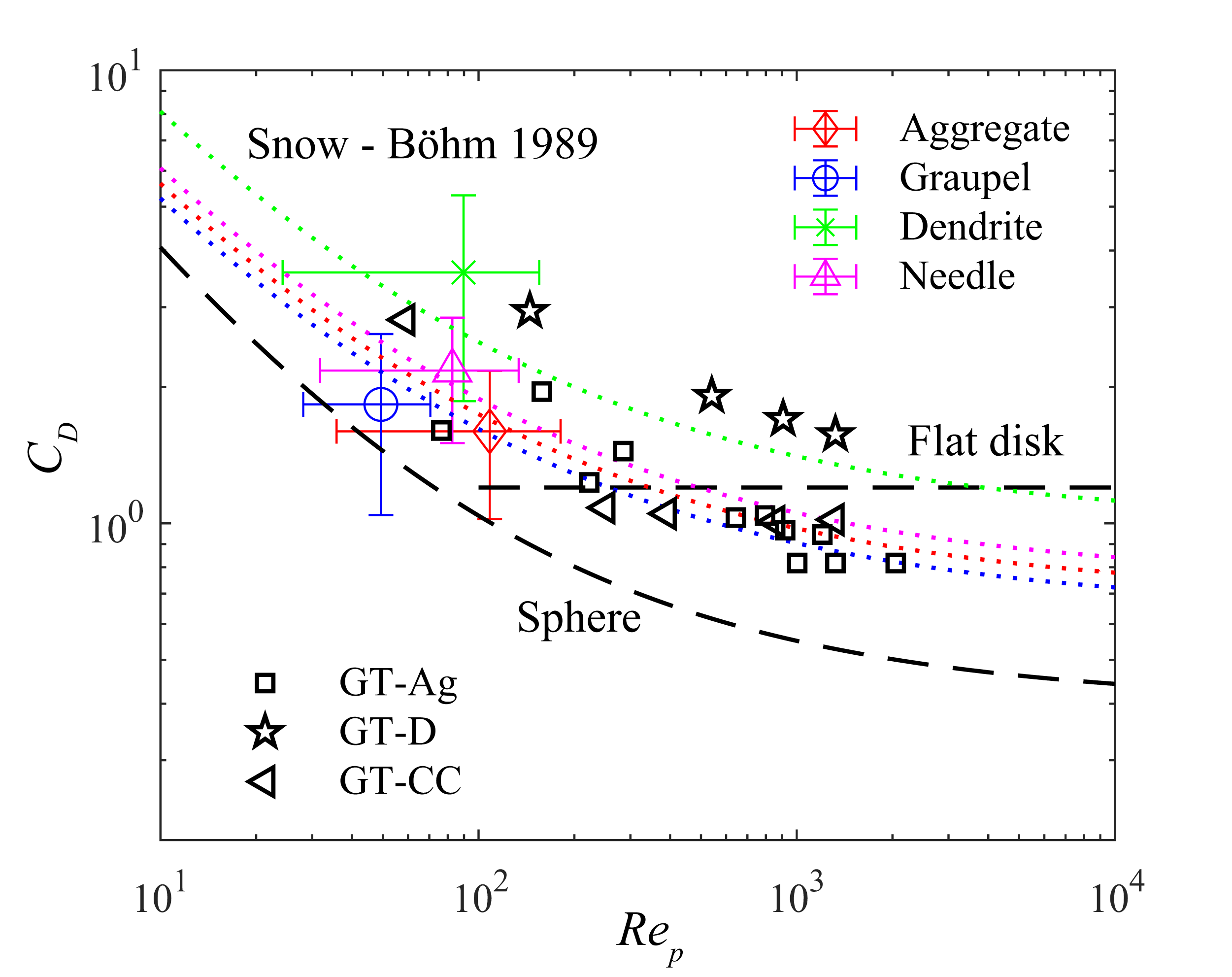}}
  \caption{The relationship between drag coefficient ($C_D$) and particle Reynolds number ($Re_p$) for four types of snow particles. Presented data points, with corresponding error bars indicating measurement uncertainties, denote mean values of $C_D$ and $Re_p$ derived from average particle size and settling velocity for each snow type: aggregates (red diamond), graupel (blue circle), dendrites (green star), and needles (magenta triangle). These measured data are compared against theoretical $C_D - Re_p$ correlations for spheres \citep[dashed line,][]{kaskas1970schwarmgeschwindigkeiten} and flat disks at high $Re_p$ values \citet{roos1971drag}, as well as with empirical correlation ($C_{De}=\left(\overline{A / A_e} \right)^{3/4} C_0\left(1+\delta_0 / R e_p^{1 / 2}\right)^2$) for natural snow particles (dotted lines, with the same color scheme as the measured drag coefficient) and recent findings from 3D-printed snow particles (black squares, pentagrams, and left-pointed triangles) by \citet{tagliavini2021drag}.}
\label{fig:6}
\end{figure}

To better model the terminal velocity, it is important to quantify the aerodynamic drag of snow particles for various morphological types. In Figure \ref{fig:6}, we present the mean drag coefficients and mean Reynolds numbers, estimated using the average particle size and measured settling velocity. The error bars indicate the variability of these quantities, reflecting the distribution of snow particle sizes and settling velocities as represented by their standard deviations. The drag coefficient is calculated as $C_{De, \text { mean }}=2 \overline{\rho_p} \overline{V_p} g / \left( \rho_a \overline{W_s^2} \overline{A_{e,\mathrm{max} }} \right)$, where $\overline{\rho_p}$ is the average snow particle density, $\overline{V_p}$ is the average particle volume \citep[different expressions for different snow particle types defined in][]{li2023snow}, $\overline{W_s^2}$ is the mean square settling velocity; and $\overline{A_{e,\mathrm{max}}}$ is the average maximal projected area of the measured snow particles (e.g. a flat falling dendrite, see Appendix \ref{appB}). The snow particles have an average Reynolds number ($R e_{p, \mathrm{ mean }}= \overline{W_s} \ \overline{D_p} / \nu$) on the order of 100, agreeing with typical field measurements \citep{heymsfield2010advances}. The drag coefficients for aggregates, graupels, and needles agree well with the model predictions from \citet[][equation \ref{eqn:2}]{bohm1989general}, $C_{De}=\left(\overline{A / A_e} \right)^{3/4} C_0\left(1+\delta_0 / R e_p^{1 / 2}\right)^2$, as presented by the dotted lines. Note that the average area ratio, $\overline{A / A_e}$, is calculated from the snow particle holograms for each snow type, and it is necessary to rescale the generalized drag equation \ref{eqn:2} to the specific snow morphologies \citep{bohm1989general}. As graupels show more sphere-like features, their drag coefficient leans towards that of spheres \citep[corrected for high Reynolds number by][]{kaskas1970schwarmgeschwindigkeiten}. Despite the smaller terminal velocity for the non-spherical particles considering the particle orientation, the drag coefficient is well-predicted by the \citet{bohm1989general} model for aggregates and needles. Potential contamination from other types ($\sim$ 20$\%$ after filtering out particles with close to an aspect ratio of one) within the datasets might lead to the mismatch between the PDFs of the terminal velocity and measured settling velocity for needles. The enhanced settling for aggregates could be a combined result of particle orientation, weak turbulence enhancement considering the critical condition of $St_\eta Sv_L \sim 1$, and contamination from other types that do not align with the statistically dominant group, as shown in Table \ref{tab:2}. Moreover, the dendrites show, on average, a higher drag coefficient as compared to the other types, potentially due to their large frontal area and higher density. Such a discrepancy could also explain the underestimation of the terminal velocity of dendrites by the equations from \citet{bohm1989general}, considering the higher, on average, settling velocity. We further compare our measurements with laboratory experiments by \citet{tagliavini2021drag}, with squares representing the aggregates (AgCr77, Ag15P1, AgSt18), pentagrams representing the dendrite (D1007), and left-pointed triangles representing the columnar snow (CC20Hex2). Our measurements agree well with the laboratory experiments, with the dendrite type showing a larger drag coefficient due to its disk-like shape. Such observations provide insights for snow settling modeling, especially for the predominantly dendrite snow events, as they show a large deviation from the \citet{bohm1989general} model prediction.

\subsection{\label{sec:32}Settling kinematics}
\subsubsection{\label{sec:321}Qualitative observation}

\begin{figure}
  \centerline{\includegraphics[scale = 0.85]{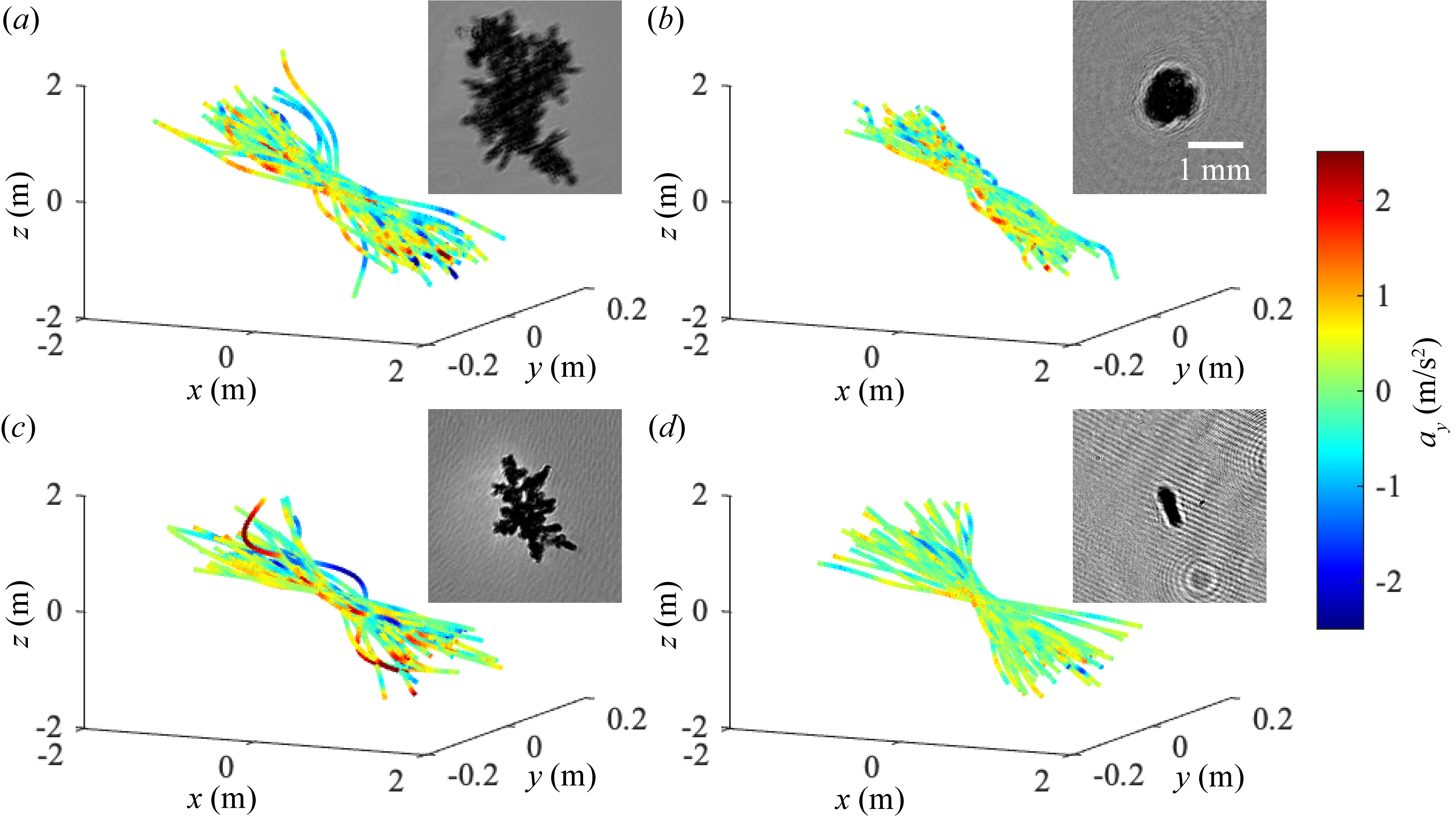}}
  \caption{A random selection of 50 trajectories for four snow particle types, with paths color-coded according to spanwise acceleration and centered at the origin. Panels (\textit{a}) aggregates, (\textit{b}) graupel, (\textit{c}) dendrite, and (\textit{d}) needles each display a group of sample trajectories. Insets provide corresponding holograms for each snow type. We remind that with our coordinate system aligned with the mean wind direction, all particles will travel towards positive \textit{x} value.}
\label{fig:7}
\end{figure}

Besides the settling velocity, the kinematic behaviors of snow particle settling trajectories are as variable as their shapes, with morphology playing a significant role in their settling behavior. Similar to the findings of Kajikawa’s laboratory studies \citep{kajikawa1976observation,kajikawa1982observation,kajikawa1989observation,kajikawa1992observations,kajikawa1997observations}, snow particles demonstrate a range of falling styles under weak atmospheric turbulence, akin to those of disks and thin cylinders in quiescent flows. Figure \ref{fig:7} displays a collection of snow particle trajectories, differentiated by the color-coded spanwise acceleration, from datasets dominated by different snow types. The distinct kinematics observed here are likely a consequence of each type’s unique morphology under similar atmospheric conditions. Aggregates and dendrites, in particular, exhibit pronounced meandering motion, characterized by substantial acceleration fluctuations at a relatively low frequency. This behavior could be attributed to their larger sizes and frontal areas, which, when subject to even weak atmospheric turbulence, result in unstable settling patterns marked by fluttering or tumbling motions. In contrast, graupels, with their quasi-spherical form, show a relatively high-frequency, low-magnitude meandering motion, and maintain a consistent travel direction. This suggests that graupels can better follow the fluid flow, considering their smaller particle size and lower density compared to the other non-spherical particle types, with their meandering motion possibly revealing interactions with small turbulent eddies. Needles exhibit weak magnitude and infrequent fluctuations in acceleration, but appear to experience a wider spanwise velocity range, as shown by the spread of trajectories in the spanwise direction. Their elongated, cylindrical shape, presenting a minimal frontal area relative to length, likely contributes to their tendency to align with the flow, resulting in this distinct settling pattern. Additionally, a detailed but qualitative examination of the trajectories indicates that non-spherical particles predominantly exhibit zig-zag motions, potentially due to the vortex shedding in their wake \citep{willmarth1964steady,toupoint2019kinematics}, whereas graupels tend to follow more helical paths, potentially spiraling around vortex tubes \citep{mezic1998regular}.

\subsubsection{\label{sec:322}Kinematic quantification}

\begin{figure}
  \centerline{\includegraphics[scale = 0.85]{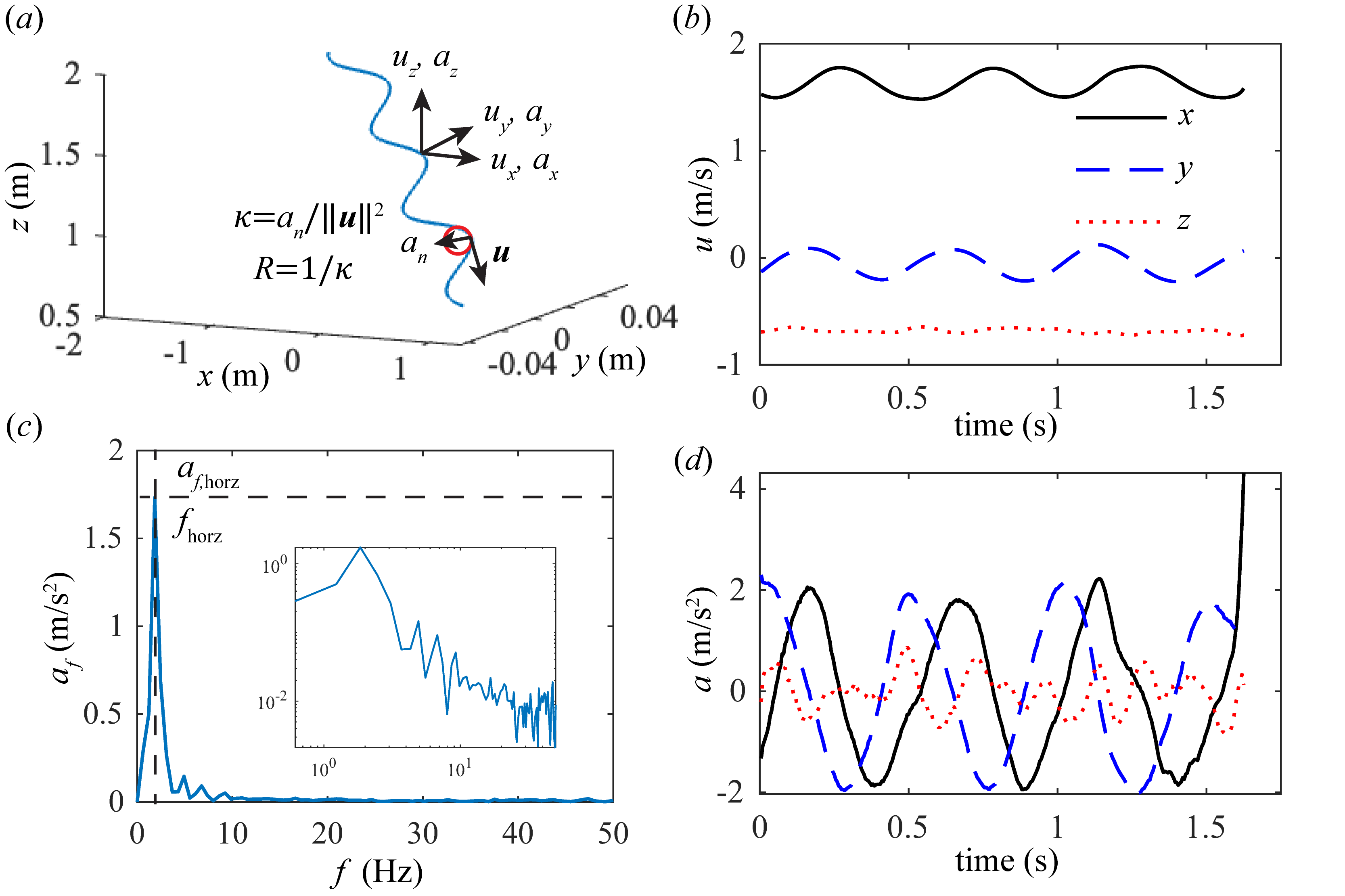}}
  \caption{Kinematic analysis of a sample trajectory. (\textit{a}) Annotated meandering trajectory of a dendrite snow particle detailing its Lagrangian velocity components $(u_x,u_y,u_z)$, accelerations $(a_x,a_y,a_z)$, curvature ($\kappa$), and radius of curvature ($R$). (\textit{b}) Temporal variations in the three velocity components along the trajectory: $x$ (black solid line), $y$ (blue dashed line), $z$ (red dotted line). (\textit{c}) Frequency spectrum of horizontal acceleration, highlighting the peak frequency and maximum fluctuation amplitude. Inset shows the same spectrum in log-log scale. (\textit{d}) Corresponding temporal variations in the three acceleration components along the trajectory with the same color scheme as in (\textit{b}).}
\label{fig:8}
\end{figure}

To thoroughly analyze the kinematics of snow particles, we examine their trajectories using the particle velocity ($\boldsymbol{u}=(u_x,u_y,u_z)$), the Lagrangian accelerations ($\boldsymbol{a}=(a_x,a_y,a_z)$), and the resulting curvature ($\kappa=\| \boldsymbol{u} \times \boldsymbol{a} \| / \|\boldsymbol{u}\|^3$), where $\times$ represents the cross product. The curvature quantifies the trajectory's deviation from a straight path, influenced by flow structures or the snow particle morphology. We define two curvatures: one based on the original path and another adjusted to reduce the effect of the different mean streamwise flow and settling velocities across datasets ($\kappa= \| \boldsymbol{u}^\prime \times \boldsymbol{a} \| / \| \boldsymbol{u}^\prime \|^3$, with $\boldsymbol{u}^{\prime}=\left(u_x-\overline{u_x}, u_y, u_z-\overline{u_z}\right)$). Figure \ref{fig:8} demonstrates this analysis with the trajectory of a dendrite snow particle. The apparent sinusoidal meandering is an actual measurement from our 3D PTV system. This meandering is underscored by the sinusoidal patterns in the velocity and acceleration components (Figure \ref{fig:8}\textit{b} and \textit{d}), particularly pronounced in the acceleration signals in the horizontal plane. Spectral analysis of the acceleration variation along specific trajectories enables us to discern the strength and frequency of the meandering motion (Figure \ref{fig:8}\textit{c}). The dominant frequency is identified from the spectral peak, and the intensity is characterized by the magnitude of the horizontal acceleration fluctuations at this frequency. This comprehensive analysis yields detailed insights into how the morphology of snow particles influences their settling dynamics, particularly highlighting the effects on their acceleration statistics and trajectory geometry, which will be explored in depth in subsequent sections.

\subsubsection{\label{sec:323}Acceleration statistics}

\begin{figure}
  \centerline{\includegraphics[scale = 0.85]{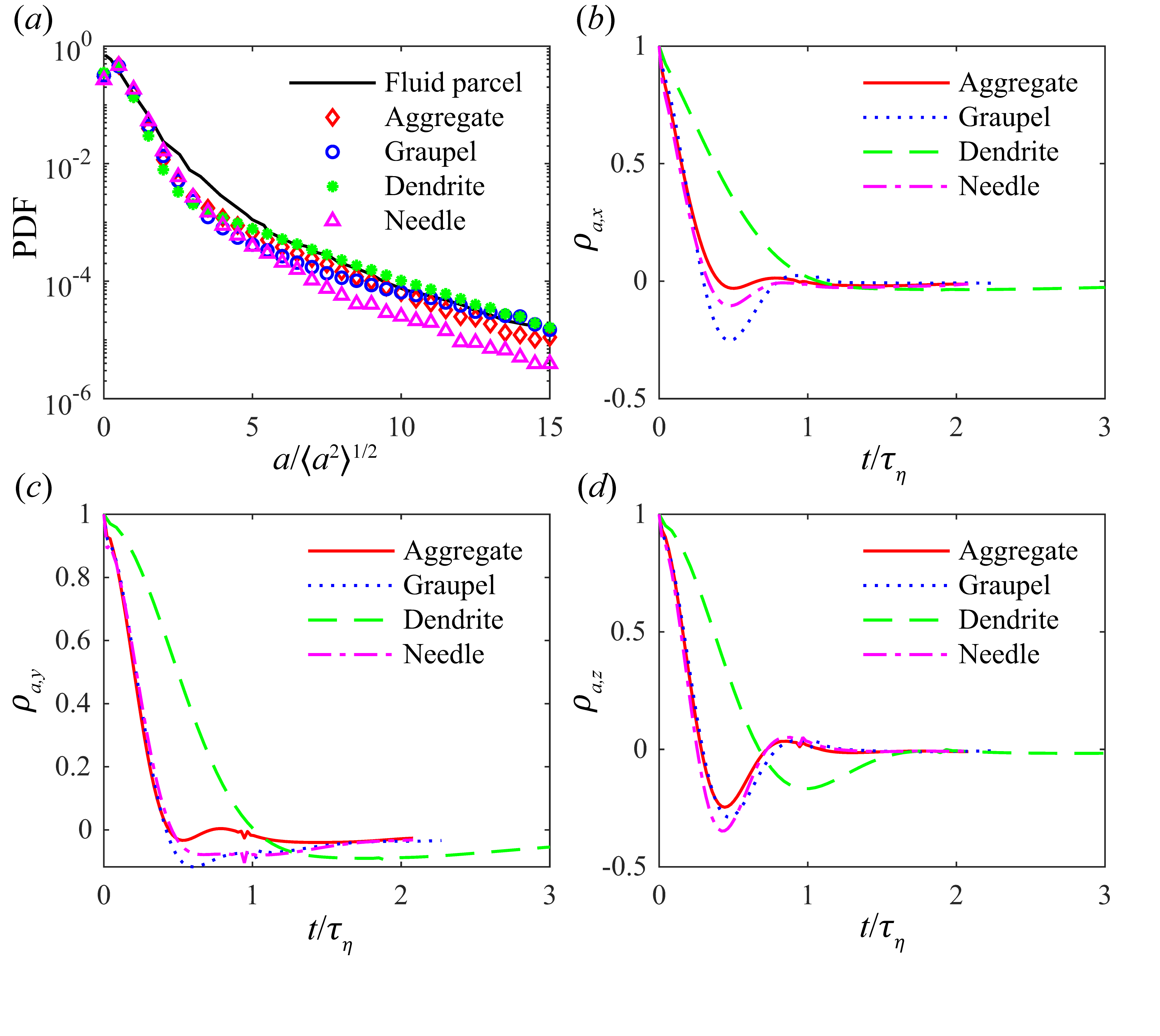}}
  \caption{(\textit{a}) Probability density functions (PDFs) of acceleration across four snow particle types —aggregates (red diamonds), graupel (blue circles), dendrites (green stars), and needles (magenta triangles)—set against the benchmark fluid parcel acceleration from homogeneous isotropic turbulence, as reported by \citet{bec2006acceleration}. (\textit{b}, \textit{c}, \textit{d}) Acceleration autocorrelation functions, (\textit{b}) $\rho_{a,x}$, (\textit{c}) $\rho_{a,y}$, and (\textit{d}) $\rho_{a,z}$, averaged from these snow particle trajectories, with each type depicted by the following color and line style: aggregates (red solid line), graupel (blue dotted line), dendrites (green dashed line), and needles (magenta dash-dotted line). The $x$-axis, temporal difference, is normalized by the Kolmogorov time scale.}
\label{fig:9}
\end{figure}

Having quantified the settling trajectories of snow particles, we proceed to examine and compare the acceleration statistics across datasets featuring four snow particle types. Figure \ref{fig:9} presents a detailed comparison of the acceleration behaviors of different types through the normalized acceleration probability density functions (PDFs) and the Lagrangian acceleration auto-correlation. The acceleration response of the particles is influenced by their morphological features and density within the weak atmospheric turbulence. Figure \ref{fig:9}\textit{a} juxtaposes the normalized acceleration PDFs of different snow types against the acceleration of fluid parcel in homogeneous isotropic turbulence, based on simulations by \citet{bec2006acceleration}. It is generally anticipated that, due to inertia, particles in turbulence will not accelerate as intensely as the surrounding fluid because they cannot keep pace with the rapid fluctuations of the turbulent flow. Nevertheless, the shape of the particles is also a critical factor in their acceleration dynamics. Dendrites, for instance, are more prone to high acceleration events, likely a consequence of their considerable size, expansive frontal area, and the non-linear nature of the drag forces they experience. Aggregates and graupels display a decrease in the probability of high accelerations, attributable to their less intricate shapes. Needles, characterized by their slender profile, exhibit a diminished probability of encountering higher acceleration events, which may be due to their streamlined shape that naturally aligns with the flow and vortex structures within, along with their smaller size and frontal area \citep{voth2017anisotropic}. The PDF tails of these non-spherical particles also appear to correlate with their shape factors, with needles being prolate ($\beta > 1$), dendrites oblate ($\beta < 1$), and aggregates displaying a spectrum in between these extremes. This observation contrasts with the recent findings reported by \citet{singh2023universal}, which provides a universal scaling for snow particle acceleration. Additional experimentation under various turbulence conditions is needed to further investigate this discrepancy. 

In Figure \ref{fig:9}\textit{b}, \textit{c}, and \textit{d}, the acceleration autocorrelation functions of the Lagrangian acceleration components reveal distinct inertial responses for the four types of snow particles. These functions are derived from the snow particles’ settling trajectories, using the formula $\rho_a(n \Delta t)= \left\langle a\left(t_0\right) a\left(t_0+n \Delta t\right)\right\rangle / \left\langle a^2\right\rangle$, where $n$ is the number of time steps, and $\Delta t=1/200$ s is the time step. Dendrites display the highest inertia, indicating a more pronounced resistance to changes in the fluid motion, followed by aggregates, needles, and graupels (small differences among the three for all components). The particle inertia is attributable to the larger sizes of dendrites and aggregates, their non-spherical shapes, and their greater density in the case of dendrites and needles. The zero-crossing points ($\tau_0$) on the autocorrelation curves also provide temporal insights into the acceleration fluctuations and, consequently, the frequency of the meandering motions of the snow particles, as listed in Table \ref{tab:4}. It scales with one-fourth the period of the meandering motion. In Table \ref{tab:4}, we present a comparison of four times the zero-crossing time (4$\tau_0$) of the acceleration autocorrelation function for the three acceleration components across various snow particle types. Generally, dendrites exhibit the largest zero-crossing time scales in their acceleration autocorrelation functions, suggesting a low-frequency meandering motion. Conversely, graupels demonstrate the smallest zero-crossing time scales, suggesting the fastest meandering frequency, corroborating the qualitative observations from Figure \ref{fig:7}. The acceleration autocorrelation functions for needles and aggregates reach their initial zero at intermediary times, with aggregates showing slightly larger time scales. The trends in these autocorrelation functions further emphasize the influence of particle morphology on settling behavior, with the aspect ratios of non-spherical particles mirroring the trends in the zero-crossing time scales.

\begin{figure}
  \centerline{\includegraphics[scale = 0.85]{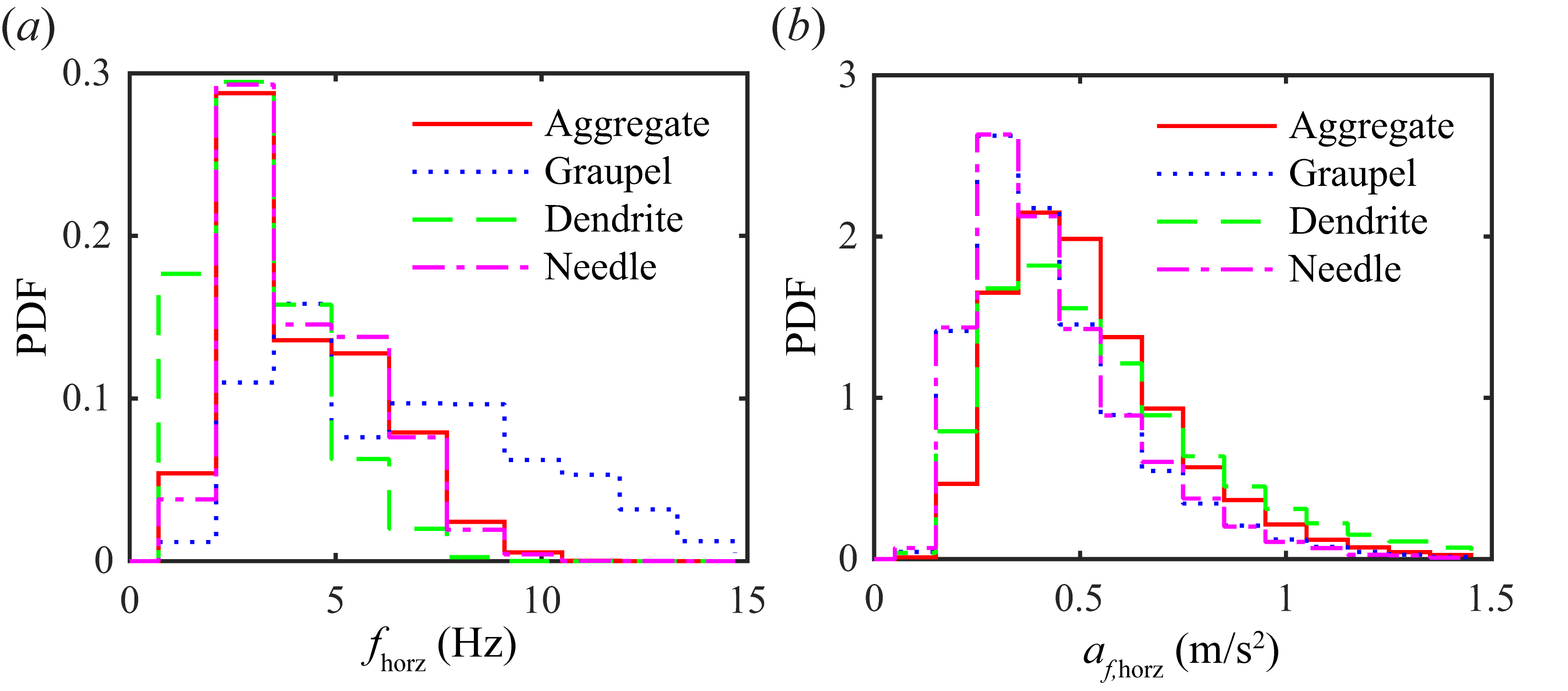}}
  \caption{(\textit{a}) Probability density functions (PDFs) of the frequency of horizontal acceleration fluctuations ($f_\mathrm{horz}$) and (\textit{b}) PDFs of the magnitude of these fluctuations ($a_{f,\mathrm{horz}}$) across four snow particle types. Aggregates are represented by red solid lines, graupels by blue dotted lines, dendrites by green dashed lines, and needles by magenta dash-dotted lines.}
\label{fig:10}
\end{figure}

Following up the autocorrelation functions of acceleration above, we provide a more direct measurement of the meandering motion of snow particles by examining the Lagrangian variations in position, velocity, and acceleration along their settling trajectories, as shown in Figure \ref{fig:8}. The horizontal acceleration component, displaying the most pronounced variation, serves as a key indicator of meandering behaviors, as illustrated in Figure \ref{fig:10}\textit{a} and \textit{b}, which depict the PDFs of acceleration fluctuation frequency and magnitude for different snow particle types. This analytical approach aligns with the qualitative findings from Figure \ref{fig:7} and supports the acceleration statistics presented in Figure \ref{fig:9}. The measured average frequencies and corresponding magnitudes are summarized in Table \ref{tab:4}. These frequencies can be nondimensionalized into Strouhal numbers, $S t=\overline{f_{\mathrm{horz }}} \cdot \overline{D_p} / \overline{W_s}$, as proposed by \citet{willmarth1964steady}, and summarized in Table \ref{tab:4}. Although the near-spherical shape of graupels is not expected to induce meandering motion, our measurements surprisingly reveal a weak meandering or helical motion, as evidenced by variations in velocity and acceleration. The observed average frequency of this motion closely matches the Kolmogorov scale frequency $1/\tau_\eta  = 4.6$ Hz. This correspondence suggests that, despite the dominance of morphological effects in dictating particle behavior, especially for non-spherical particles, graupels still move around and weakly interact with the Kolmogorov eddies within the flow, considering their sizes close to those of the Kolmogorov eddies.  In contrast, the meandering frequencies for non-spherical particles are lower than both the frequency corresponding to the Kolmogorov scale and the vortex shedding frequency in the wake of anisotropic particles identified in various studies \citep{willmarth1964steady,auguste2013falling,tinklenberg2023thin,jayaweera1965behaviour,toloui2014measurement}. Specifically for dendrites, we estimate the dimensionless moment of inertia to be $\sim O(0.1-1)$, resulting in the Strouhal number $\sim O(0.01)$ based on Willmarth et al. (1964), larger than that of the dendrites from our measurement. This discrepancy may be attributable to the delayed inertial response of non-spherical particles to the fluid flow and vortex shedding, as well as to the permeability of the dendrites. Moreover, the measured Strouhal numbers for these particles are consistent with Kajikawa’s laboratory measurements \citep{kajikawa1976observation,kajikawa1982observation,kajikawa1989observation,kajikawa1992observations,kajikawa1997observations}, situating our findings within the observed range for the meandering motions of non-spherical snow particles. The vertical acceleration component displays fluctuations that could stem from orientation changes (resulting in drag force variation) in anisotropic particles due to horizontal meandering. The $a_z$ fluctuation magnitudes are more pronounced since the horizontal component combines the $x$ and $y$ components, which are typically out of phase. Furthermore, the vertical acceleration fluctuation frequency is nearly twice the horizontal one because the inferred changes in particle orientation, caused by horizontal meandering, e.g., a perfectly edge-on configuration, have a 180-degree periodicity for disk- and needle-like shapes. This interpretation is consistent with the minimal differences observed for the near-symmetric graupel, and the trend is also consistent with the shorter zero-crossing times ($\tau_{0,z}$) observed in our data.

\begin{table}
  \begin{center}
\def~{\hphantom{0}}
    \begin{tabular}{|c|c|c|c|c|c|c|c|c|c|}
\hline    \begin{tabular}{c} Snow type \end{tabular} & \begin{tabular}{c} $\overline{a_{f, \mathrm{ horz }}}$ \\ m $/ \mathrm{s}^2$ \end{tabular} & \begin{tabular}{c} $\overline{a_{f, z}}$ \\ m $/ \mathrm{s}^2$ \end{tabular} & \begin{tabular}{c} $\overline{f_{\mathrm{horz}}}$ \\ $\mathrm{Hz}$ \end{tabular} & \begin{tabular}{c} $\bar{f}_z$ \\ $\mathrm{Hz}$ \end{tabular} & \begin{tabular}{c} $S t=\overline{f_{\text {horz }}} \cdot \overline{D_p} / \overline{W_s}$ \\ - \end{tabular} & \begin{tabular}{c} $4 \tau_{0, x}$ \\ $\mathrm{s}$ \end{tabular} & \begin{tabular}{c} $4 \tau_{0, y}$ \\ $\mathrm{s}$ \end{tabular} & \begin{tabular}{c} $4 \tau_{0, z}$ \\ $\mathrm{s}$ \end{tabular} & \begin{tabular}{c} $\tau_\eta$ \\ $\mathrm{s}$ \end{tabular} \\
\hline    Aggregate & 0.27 & 0.60 & 2.6 & 3.8 & $2 \times 10^{-3}$ & 0.38 & 0.42 & 0.27 & 0.24 \\
\hline    Graupel & 0.19 & 0.45 & 5.6 & 6.6 & $4 \times 10^{-3}$ & 0.26 & 0.36 & 0.26 & 0.22 \\
\hline    Dendrite & 0.30 & 0.47 & 2.0 & 3.3 & $2 \times 10^{-3}$ & 0.51 & 0.48 & 0.32 & 0.12 \\
\hline    Needle & 0.22 & 0.55 & 2.5 & 4.1 & $1 \times 10^{-3}$ & 0.32 & 0.44 & 0.24 & 0.24 \\ \hline
    \end{tabular}
  \caption{Comparative summary of horizontal and vertical acceleration variations in snow particle trajectories, presenting the average magnitude ($\overline{a_{f,\mathrm{horz}}}$, $\overline{a_{f,z}}$) and frequency ($\overline{f_\mathrm{horz}}$, $\overline{f_z}$), alongside the zero-crossing times ($\tau_{0,x}$, $\tau_{0,y}$, and $\tau_{0,z}$) of the acceleration auto-correlation functions and the Kolmogorov time scale ($\tau_\eta$), for different snow particle types.}
  \label{tab:4}
  \end{center}
\end{table}

\subsubsection{\label{sec:324}Trajectory geometry}

\begin{figure}
  \centerline{\includegraphics[scale = 0.8]{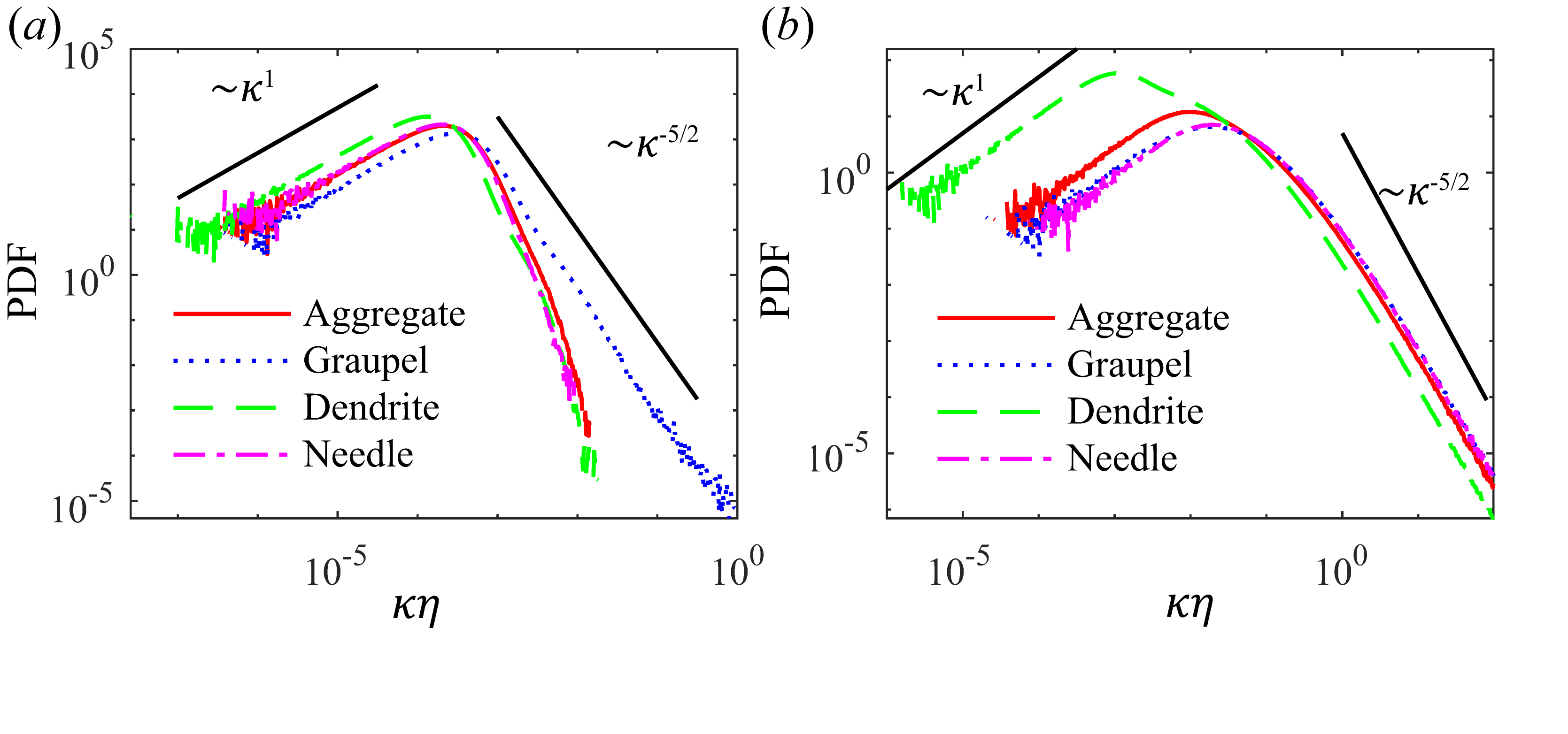}}
  \caption{(\textit{a}) Probability density functions (PDFs) of the normalized trajectory curvature ($\kappa\eta$, normalized by the Kolmogorov scale) for four different snow particle types using original path data. (\textit{b}) PDFs of normalized curvature after adjusting for the mean streamwise flow and settling velocities. Each snow type is depicted by a distinctive line style and color: aggregates with red solid lines, graupel with blue dotted lines, dendrites with green dashed lines, and needles with magenta dash-dotted lines.}
\label{fig:11}
\end{figure}

The variance in meandering frequency and magnitude across different types of snow particles results in distinctive trajectories. To quantify their geometrical differences, we employ curvature calculations both with and without the impact of the mean streamwise and settling velocities. Figure \ref{fig:11} illustrates the probability distribution functions (PDFs) of these normalized trajectory curvatures ($\kappa\eta$, where $\eta$ is the Kolmogorov scale). For the original trajectories, curvature is calculated using the formula $\kappa= \| \boldsymbol{u} \times \boldsymbol{a} \| / \|\boldsymbol{u}\|^3$ (Figure \ref{fig:11}\textit{a}), where $\times$ indicates the cross product between the velocity ($\boldsymbol{u}$) and acceleration ($\boldsymbol{a}$) vectors. Additionally, to minimize the influence of varying flow and settling velocities across datasets, we adjust the velocity vector to $\boldsymbol{u}^{\prime}=\left(u_x-\overline{u_x}, u_y, u_z-\overline{u_z}\right)$), and recompute curvature (Figure \ref{fig:11}\textit{b}). Previous research by \citet{braun2006geometry}, \citet{xu2007curvature}, and \citet{scagliarini2011geometric} has explored the geometry of fluid trajectories in turbulence, uncovering characteristic scaling within the curvature PDFs. Their findings suggest a universal scaling for both tails of the PDFs: low curvature events scale with $\kappa^1$, while high curvature events follow a $\kappa^{-5/2}$ scaling. \citet{xu2007curvature} propose that these tail scaling laws result from Gaussian velocity statistics rather than turbulence gradients, contending that high curvature events correlate with periods of low velocity rather than high acceleration from interactions with thin vortex tubes as one might expect. Moreover, curvature can also be expressed as $\kappa= \| a_n \| / \|\boldsymbol{u}\|^2$, so the tail of the curvature PDF, $P_{\kappa \rightarrow \infty}$, as $\kappa \rightarrow \infty$, scales similarly to the tail of the PDF of $\boldsymbol{u}^{-2}=1 / \left( u_x^2+u_y^2+u_z^2 \right)$, $P_{\boldsymbol{u}^{-2} \rightarrow \infty}$, as $\boldsymbol{u}^{-2} \rightarrow \infty$. Assuming velocity components are independent and follow Gaussian statistics, $P_{\boldsymbol{u}^{-2} \rightarrow \infty}$ conforms to a chi-square distribution with three degrees of freedom, leading to the derived scaling of $P_{\kappa \rightarrow \infty} \sim \kappa^{-5/2}$. \citet{bhatnagar2016deviation} extended this theoretical framework to heavy inertial particles and verified through simulations that the same scaling applies to the PDFs of these particles’ trajectories. These theoretical insights can be integrated into our analysis of the geometry of snow particle trajectories, providing a better understanding of the intricate settling dynamics and trajectory geometry under the influential role of particle morphology. 

Figure \ref{fig:11}\textit{a} and \textit{b} reveal that for most snow particle types, the tails of the curvature probability distribution functions (PDFs) exhibit similar scaling trends as reported in the previous research  \citep{braun2006geometry,xu2007curvature,scagliarini2011geometric,bhatnagar2016deviation}. Nonetheless, when considering the mean streamwise and settling motions, the PDFs of curvature for non-spherical particles exhibit a notably different scaling, approximately following a $\kappa^{-4}$ trend as shown in Figure \ref{fig:11}\textit{a}. This deviation may arise from the rotation and meandering motion due to the morphology of non-spherical snow particles, which modulates the Lagrangian velocities along the trajectories. Notably, when the mean settling and streamwise velocities are removed from consideration, the tails of the curvature PDFs for different snow types tend to align on the higher curvature end. This pattern indicates that particle morphology predominantly influences the mean values of the settling and streamwise velocities, rather than their fluctuations. Moreover, the peaks of the PDFs are around $10^{-2}$ and $10^{-3}$, similar to those in the previous studies. However, as proposed by \citep{xu2007curvature}, for fluid tracers, the peak of the PDF scale with $\left(\eta R e_\lambda\right)^{-1}$, which for the snow particle trajectories would be $\sim O(1)$. The smaller curvature for the snow particle trajectories might be attributed to the particle inertia \citep{maxey1987gravitational}. Further analysis shows that, despite the pronounced meandering behavior of dendrites, they exhibit the smallest mean curvature, with aggregates, needles, and graupels following in ascending order. This trend can be explained by the fact that both the frequency and magnitude of the meandering motion contribute to the overall trajectory curvature. Dendrite trajectories, while displaying significant fluctuations in spanwise meandering motion, have a lower frequency, which culminates in a reduced mean curvature. The observed differences in the curvature probability density functions (PDFs) for graupels and other non-spherical snow types can be elucidated by drawing upon our earlier analysis in Section \ref{sec:323}. For graupels, the curvature PDF in Figure \ref{fig:11}\textit{a} scales like that of fluid trajectories, indicating that the weak meandering behavior of graupels may stem from interactions with turbulent eddies. In contrast, for non-spherical particles, it is likely due to the combined influence of wake vortex instabilities, as discussed in Section \ref{sec:323}, and weak atmospheric turbulence, as the scaling later converged in Figure \ref{fig:11}\textit{b}. 

\begin{figure}
  \centerline{\includegraphics[scale = 0.8]{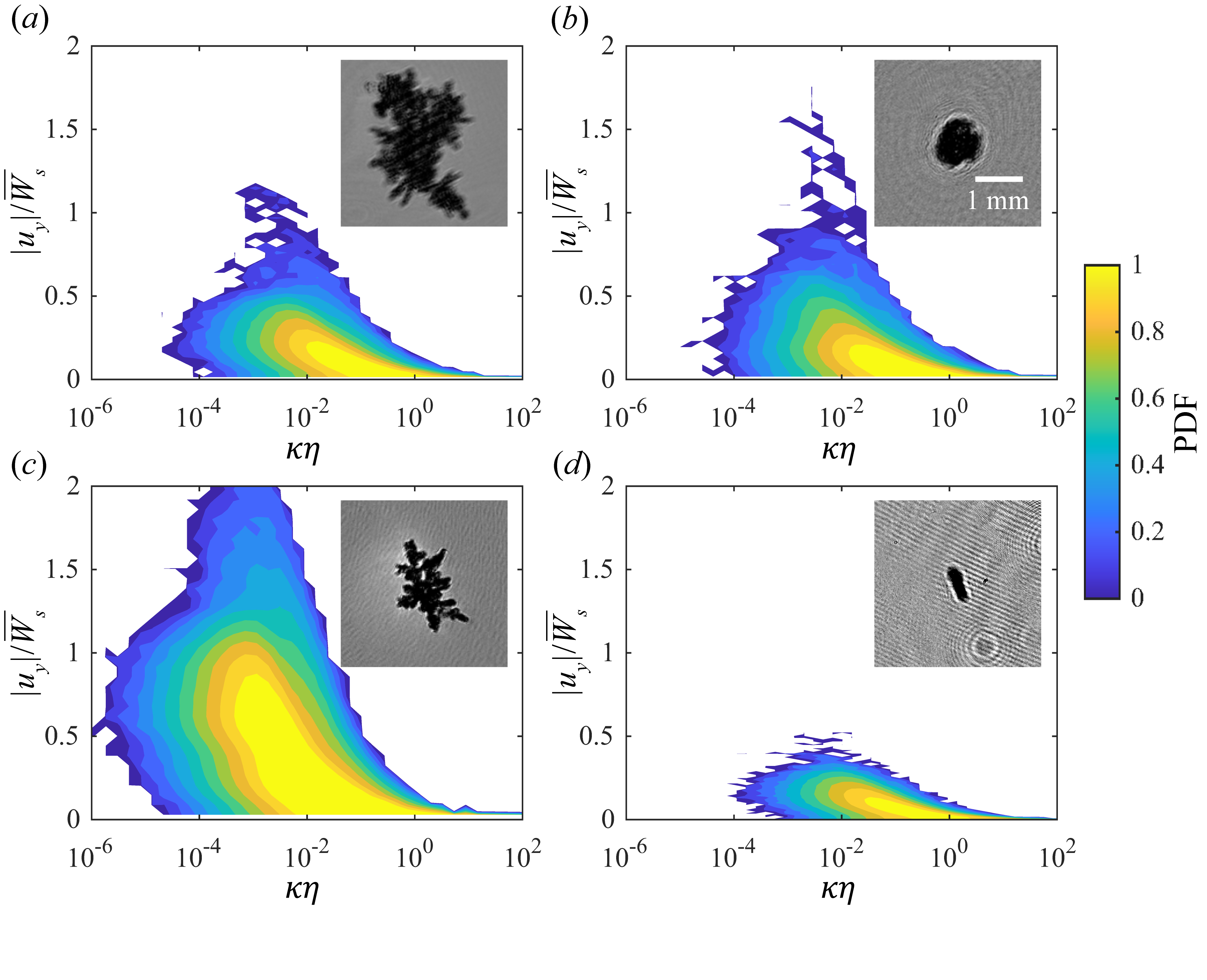}}
  \caption{Joint probability density functions (PDFs) depicting the interdependence of normalized spanwise velocity magnitude ($\left|u_y\right| / \overline{W_s}$) and normalized trajectory curvature ($\kappa\eta$) for four types of snow particles: (\textit{a}) aggregates, (\textit{b}) graupel, (\textit{c}) dendrites, and (\textit{d}) needles. The color gradient indicates the probability of data occurrence, with warmer colors representing higher concentrations. Insets provide sample holograms of typical particles from each type.}
\label{fig:12}
\end{figure}

Figure \ref{fig:12} then delves into the relationship between normalized spanwise velocity ($\left|u_y\right| / \overline{W_s}$) and normalized trajectory curvature ($\kappa\eta$) for snow particles, taking into account the theoretical finding by \citet{xu2007curvature} that high curvature events tend to coincide with low velocities. While snow particles settling in the atmosphere generally have non-zero streamwise and settling velocities, high curvature events are often tied to moments when the spanwise velocity is minimal and changing sign. This trend is evident in Figures \ref{fig:7} and \ref{fig:8}\textit{a}, where the spanwise velocity approaches zero and reverses direction at the peaks of the meandering motion, leading to increased curvature at these turning points. In Figure \ref{fig:12}, joint PDFs map the spanwise velocity magnitude and the local trajectory curvature, after subtracting the mean streamwise and settling velocities, for each snow particle type. A pronounced negative correlation between spanwise velocity and trajectory curvature is observed, particularly for dendrites (Figure \ref{fig:12}\textit{c}), which exhibit the most substantial correlation coefficient ($\sigma_{xy}=-0.80$). Aggregates display a similar negative correlation, but with a slightly lower coefficient ($\sigma_{xy}=-0.77$) and a reduced magnitude of spanwise velocity. Needles, despite having the lowest spanwise velocity magnitude potentially due to their smaller frontal area and high density, maintain a strong correlation with curvature, indicated by a correlation coefficient of $\sigma_{xy}=-0.76$. Graupels, on the other hand, show the weakest correlation among all particle types, with the lowest coefficient ($\sigma_{xy}=-0.71$). This analysis highlights the profound effect of snow particle morphology on meandering motion for non-spherical particles, which associates the near-zero spanwise velocity in the meandering extremes with the high curvature in their trajectories. Unlike graupels, whose weaker meandering motion is influenced more by interactions with turbulence eddies, non-spherical particles do not exhibit the expected correlation between high curvature and high acceleration, suggesting that their complex morphology dominates this meandering motion and corresponding high curvature events.

\subsection{\label{sec:33}Interconnection between trajectory geometry and settling velocity}

Our comprehensive analysis elucidates the distinctive settling behaviors of snow particles with various morphologies, addressing the questions raised at the beginning of our results section. Concerning the influence of morphology on snow aerodynamic properties, we observe that the response times for all snow particles are broadly similar, averaging around 0.1 seconds, on the same order of the intercept in the acceleration autocorrelation function. However, needles exhibit a marginally increased response time attributed to their higher density, particularly when compared with the density of the surrounding air. This higher density contributes to the needles' higher average terminal velocity in still air. Furthermore, the empirical models by \citet{bohm1989general} well predict the drag coefficients for aggregates, graupels, and needles, while dendrites emerge as anomalies, exhibiting drag coefficients exceeding model predictions, likely due to their large frontal area and oblate, disk-like, geometry. Notably, although dendrites and needles have similar aspect ratios, as measured by the snow particle analyzer, the dendrites are disk-like, oblate spheroids, while the needles are columnar, prolate spheroids. 

The different aerodynamic properties and morphologies of these snow particles have strong effects on their settling kinematics. Specifically, dendrites display unique behavior compared to other types. Their non-linear drag and substantial frontal area result in the most prominent acceleration fluctuation magnitude, occurring at relatively low frequencies. While the acceleration probability density function (PDF) for dendrites closely resembles that of a fluid parcel in turbulence, the acceleration auto-correlation function indicates a slow response to the rapid fluctuation of the flow velocity. Conversely, needles exhibit minimal acceleration fluctuation magnitude, suggesting relatively large inertia and a tendency to avoid intense cross-flow drag and convoluted trajectories. Yet, the acceleration autocorrelation function indicates a moderate rate of change in the direction of acceleration. Such different behaviors from that of dendrites are possibly due to their streamlined shape aligning with fluid flow structures, as for fibers in turbulence. Overall, acceleration statistics appear to correlate with the shape factors of these particles considered as spheroids, with dendrites and needles representing the spectrum's extremes and aggregates positioned in between. 

Finally, to answer the third question, it becomes apparent, from our analysis above, that the combination of turbulence and non-spherical particle morphologies can modulate the particle settling velocities even under weak atmospheric turbulence. For the cases investigated here, we hypothesize that dendrites exhibit an enhanced settling velocity (Figure \ref{fig:5}\textit{c}) that is due to an underestimation of the drag coefficient in the model by \citet{bohm1989general}; graupels settling velocity is well predicted even though spherical particles smaller than the Kolmogorov scale close to critical Stokes conditions were expected to exhibit settling velocity enhancement. The significant cross-flow velocities (considering the large settling parameter $Sv_L$) experienced by graupels may suggest that preferential sweeping was not the only mechanism in play during settling. Aggregates drag coefficient is well captured in the still air model, implying that the observed enhanced settling is likely due to combined effects of anisotropic particle orientation and the weak atmospheric turbulence. The observed conditions are marked by $St_\eta Sv_L \sim 1$ for which settling enhancement has been predicted and observed \citep{petersen2019experimental,brandt2022particle}. Disentangling turbulence and morphology effects is challenging because turbulence-induced disturbances alter the preferential orientation (i.e., particles with their largest projected area facing the settling direction) of stably falling particles. The meandering motions of the non-spherical particles, whether fluttering or tumbling, likely affect their orientation, reducing their average projected area compared to a steady settling and thus enhancing settling velocity. However, direct measurement of particle orientation during settling is technically challenging and beyond our current capability. 

\begin{figure}
  \centerline{\includegraphics[scale = 0.8]{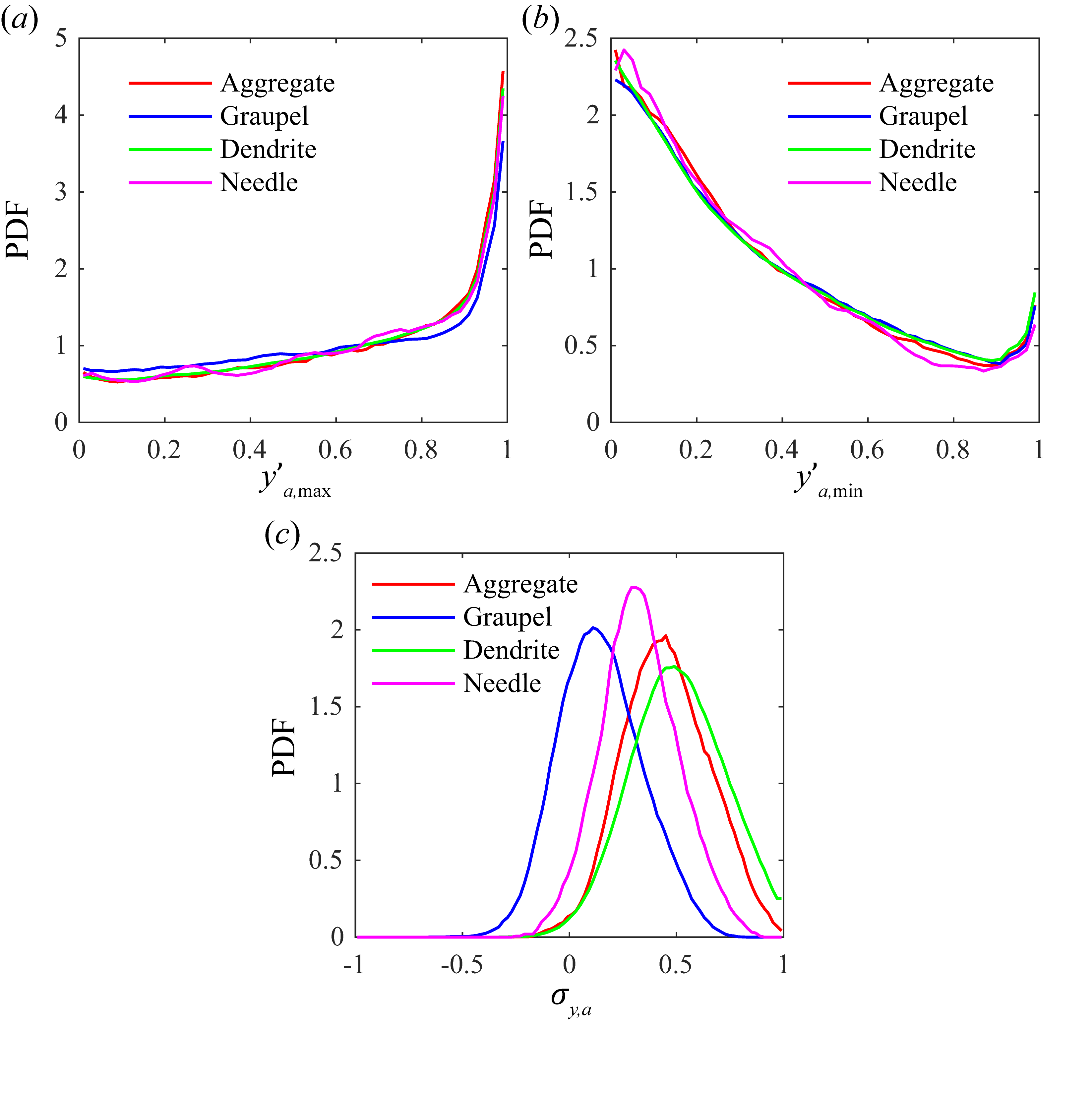}}
  \caption{(\textit{a},\textit{b}) Probability density functions (PDFs) of the spanwise snow particle positions, normalized by its maximum excursion for each trajectory, conditioned on high ($y^\prime_{a,\mathrm{max}}$) and low ($y^\prime_{a,\mathrm{min}}$) values of the vertical accelerations, and categorized by four snow particle types: aggregates (red), graupels (blue), dendrites (green), and needles (magenta). (\textit{c}) PDFs of the correlation coefficients ($\sigma_{y,a}$) between the normalized spanwise position and the downward (positive) acceleration for these particles throughout their meandering path.  Note that the spanwise particle locations $y(t)$ and corresponding vertical accelerations $a_z (t)$ are slightly temporally shifted to account for the response of the particle acceleration to change in orientation.}
\label{fig:13}
\end{figure}

Thus, we investigate the interconnection between the meandering motion and the vertical acceleration along the trajectories of snow particles. Recognizing that these fluctuations might not be perfectly synchronized (owing to the particles' inertial response and the variability in drag force related to changes in projected area and settling velocity), we have considered a slight phase shift between the varying spanwise location $y(t)$ and vertical acceleration $a_z (t+\tau)$ along the trajectories. This adjustment aims to align the locations of maximum meandering with the smallest projected area, which typically corresponds to greater downward acceleration. To maintain the integrity of the correlation, we limit the phase shift to 0.15 times a quarter of the meandering period to avoid creating an inverse relationship between vertical acceleration and meandering motion. Note that this time lag ($\tau$) is on the order of $0.1\tau_p$, and close to the estimated Strouhal number for the disk-like particles. Thus, during a fraction of the anisotropic particle rotation and the corresponding translational response time, the particle is experiencing a reduction in drag area, leading to its acceleration downward.  Figure \ref{fig:13}\textit{a} and \textit{b} demonstrate that, following this phase shift, the greatest downward accelerations predominantly occur at the furthest extent of the spanwise meandering motion ($y^\prime_{a,\mathrm{max}} \sim 1$). Conversely, the least downward acceleration—or even upward acceleration—tends to happen near the central position ($y^\prime_{a,\mathrm{min}} \sim 0$), where anisotropic snow particles are likely to have their maximum projected area facing downwards. Subsequently, we calculate the correlation coefficient between vertical acceleration and spanwise position during the snow particles’ meandering motion. The results reveal substantial positive correlation coefficients for dendrites ($\overline{\sigma_{y,a}}=0.58$) and aggregates ($\overline{\sigma_{y,a}}=0.45$), which is consistent with their observed enhanced settling velocities. Needles display a moderate correlation coefficient ($\overline{\sigma_{y,a}}=0.33$), reflective of their anisotropic shape. Graupels, however, exhibit a low average correlation coefficient of $\overline{\sigma_{y,a}}=0.15$, as changes in particle orientation are not expected to significantly affect the drag force. 

\section{\label{sec:cd}Conclusions and discussion}

In this study, we conduct a comprehensive field investigation into the three-dimensional (3D) settling dynamics of snow particles under weak atmospheric turbulence. This investigation was enabled by a field 3D particle tracking velocimetry (3D PTV) system \citep{bristow2023imaging}, recording over a million settling trajectories for four distinct types of snow particles (i.e., aggregates, graupels, dendrites, needles) and by simultaneous characterization of their aerodynamic properties using a holographic snow particle analyzer \citep{li2023snow}. We have examined the snow particle aerodynamic properties, including terminal velocity in air, settling velocity, drag coefficient, settling kinematics including acceleration statistics and trajectory geometry, and the interconnection between observed the meandering path and settling velocity of the snow particles. The comparison between the estimated terminal velocity \citep{bohm1989general} and the measured settling velocity demonstrate that non-spherical particles, especially aggregates and dendrites, exhibit large differences between measurements and model predictions potentially due to dynamic orientation changes along their meandering paths, which is not observed in graupels. Specifically, the settling enhancement observed in aggregates is likely a synergistic result between morphology-induced oscillations due to vortex shedding and the ambient flow that promotes wake instabilities and the onset of meandering motions. Even though dendrites are characterized by a higher drag coefficient, as compared to other snow types, corroborating the laboratory findings of \citet{tagliavini2021drag}, their settling velocity under weak atmospheric turbulence is higher than model predictions assuming nominal flat-falling drag area \citep{bohm1989general}. These apparently contradicting results emphasize the need to quantify particle settling dynamics along their complex trajectories in the field. A detailed Lagrangian analysis reveals that dendrites and aggregates undergo pronounced meandering motions in the horizontal plane perpendicular to the direction of gravity at relatively lower frequencies, likely governed by their inertia to tumbling and rotation but enabled by ambient turbulence. Needles, however, exhibit weaker meandering amplitudes due to their smaller frontal area. Graupels, despite their near-spherical form, undergo oscillatory motions along their settling paths, characterized by higher frequencies comparable to the Kolmogorov scale and smaller amplitudes. This behavior suggests a limited interaction with small-scale turbulence structures under the conditions investigated, a notion corroborated by the agreement between model-predicted terminal velocities and measured settling velocities. These distinct settling motions and styles are also reflected in the curvature statistics, differentiating non-spherical particles from graupels. More specifically, the analysis of vertical acceleration during meandering paths reveals that periodic changes in the orientation of non-spherical particles, especially dendrites and aggregates, contribute to their enhanced settling velocity. These findings highlight the dominant impact of the morphology of snow particles on their settling dynamics under weak atmospheric turbulence. 

Our current study provides a unique dataset of realistic snow morphologies and settling trajectories. These measurements contribute to the modeling and simulation of snow settling velocity and subsequent snow accumulation rate on the ground. Despite the dominant morphology effect, interactions between the snow particles and the weak atmospheric turbulence are still manifested in some aspects of their settling dynamics. Although the non-spherical particles are likely to rotate or tumble when settling in quiescent flow, disturbances by the ambient flow promote these unsteady motions. Besides, the meandering motion of graupels exhibits frequencies closest to that of the Kolmogorov scale, hinting at weak interaction with the ambient turbulence. The weak turbulence effect is also manifested in the curvature statistics of the particle trajectories. We observe distinct scaling laws for the high-curvature tails of the probability density functions (PDFs), which mark differences between spherical particles, consistent with \citet{xu2007curvature} for fluid trajectories in turbulence and \citet{bhatnagar2016deviation} for inertial particles, and non-spherical particles, hinting at morphological influences on their settling kinematics. To compensate for variations in streamwise flow velocity and settling velocity across different datasets and morphologies, we have corrected the curvature formulation. The resulting high-curvature tails of the PDFs converge to a universal scaling associated with low spanwise velocity, reinforcing the concept that high curvature events are associated with the meandering motions of non-spherical particles. This association is further emphasized by the observed inverse correlation between the trajectory curvature and spanwise velocity. Our detailed characterization of the snow particle morphology, density, and settling velocity may lead to an improved prediction of ground snow accumulation, benefiting several related applications in snow hazard warning, climate modeling, and traffic regulation during/after snowfall.

Despite our major findings that substantiate the hypothesis of strong morphological effects on dictating snow particle settling dynamics under conditions of weak atmospheric turbulence, several challenges persist. First, quantifying the exact enhancement or hindrance of settling velocities due to weak atmospheric turbulence remains a challenging task. A better model that elaborates on the interplay between particle morphology and turbulence will be necessary. Second, current predictive models, including those by \citet{bohm1989general}, fall short in estimating the aerodynamic properties, especially the drag coefficient, for dendrites. There is a clear need for refined models that can more accurately represent these unique and complex snow particle types. Third, while we aimed to correlate the meandering motion and orientation changes to enhanced settling in non-spherical particles, the spatial resolution of our 3D PTV system is insufficient for capturing the orientation dynamics of particles throughout their settling. Some of the smaller particles captured by the snow particle analyzer might also exhibit too weak a signal to be detected by the 3D PTV system. Advancements in measurement systems could enable simultaneous assessments of particle orientation and settling trajectory and higher resolution for capturing smaller snow particles. Systems such as a high-magnification, high-resolution 3D PTV \citep{marcus2014measurements,leinonen2021reconstruction}, or a digital inline holography setup with an expanded field of view \citep{wu2015simultaneous,li2023snow} and higher sampling rate, hold promise for the desired measurements. Finally, the variability of field conditions presents an additional layer of complexity. Snow particle types and concentrations, as well as average wind speed and direction, are subject to change over each measurement period, which typically spans 3-5 hours. The relatively slow streamwise wind adds to the difficulty of accurately estimating turbulence quantities. Looking ahead, we aim to extend our investigations to scenarios involving moderate to intense turbulence. By contrasting the settling behaviors across a spectrum of turbulence intensities, particularly for different snow particle morphologies, we anticipate a more thorough understanding of how turbulence and snow particle morphology collectively influence the settling dynamics of snow particles. This future research will enable us to improve our predictive capabilities of snow settling velocity and, in the long term, of the spatial distribution and intensity of snow accumulation on the ground during snowfalls.


\appendix
\section{}\label{appA}
\begin{figure}
  \centerline{\includegraphics[scale = 1]{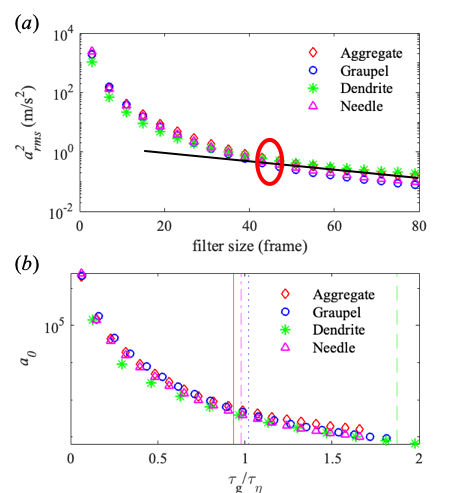}}
  \caption{(\textit{a}) The variation of the mean square of the acceleration fluctuation with the increasing Gaussian filter length. The black line represents the exponential fit, and the red circle identifies the optimal filter length. The red diamonds represent aggregates, blue circles for graupels, green stars for dendrites, and magenta triangles for needles. (\textit{b}) The normalized acceleration variance ($a_0=\left\langle a^{\prime 2}\right\rangle v^{1 / 2} / \varepsilon^{3 / 2}$) as a function of the filter length ($\tau_g$) normalized by the Kolmogorov time scale ($\tau_\eta$). The vertical lines indicate the selected filter length for different snow particle types.}
\label{fig:14}
\end{figure}

We employ a Gaussian kernel as a low-pass filter to reduce uncertainties in determining the 3D positions of snow particles and to prevent these errors from affecting Lagrangian statistics. Large errors typically occur at the trajectory ends after filtering; therefore, these segments are excluded from the statistical analysis. The selection of the kernel size is critical. A kernel that is too short fails to sufficiently reduce position uncertainty, while a kernel that is too long may suppress genuine strong acceleration events. We optimize the kernel size by analyzing the change in acceleration variance, defined as $a_0=\left\langle a^{\prime 2}\right\rangle v^{1 / 2} / \varepsilon^{3 / 2}$, across varying kernel sizes. As shown in Figure \ref{fig:14}, this approach enables the identification of the optimal kernel size. We determined that a minimal kernel size of 45 frames best maintains the exponential dependency of acceleration variance on kernel size. Notably, this selected kernel size, $\tau_g$, is comparable to the Kolmogorov length scale, $\tau_\eta$, corroborating findings from previous studies \citep{voth2002measurement, gerashchenko2008lagrangian, nemes2017snowflakes}. The estimated uncertainty in the acceleration measurements reflects the uncertainty in the filter size, which ranges between 43 and 47 frames. This results in the root mean square error in acceleration estimation, $a_{rms}$, ranging between 0.32 and 0.38 m/s$^2$ for different snow particle types.

\section{}\label{appB}
\begin{figure}
  \centerline{\includegraphics[scale = 1]{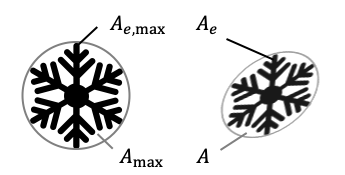}}
  \caption{The illustration on the left defines the maximum projected area ($A_{e,\mathrm{max}}$) and the maximum circumscribed area ($A_\mathrm{max}$) of a dendrite snow particle when oriented downward. The illustration on the right defines the general projected area ($A_e$) and the circumscribed area ($A$) of a dendrite snow particle in any orientation.}
\label{fig:15}
\end{figure}

To more accurately model the drag coefficient of snow particles, significant efforts have been made by researchers \citep{bohm1989general, heymsfield2010advances,mccorquodale_trail_2021}. The illustrations in Figure \ref{fig:15} summarize and clarify the calculations used in this study. The drag coefficient of snow particles can be defined using either the projected area, $C_{De}=f(A_e)$, or the circumscribed area, $C_D=f(A)$. According to \citet{bohm1989general}, the two drag coefficients are correlated by the area ratio, $C_{De} / C_D = \left(A / A_e\right)^{3/4}$, where $C_D$ is defined by equation \ref{eqn:2}. Thus, to compare the model ($C_{De} = \left(A / A_e\right)^{3/4} C_0\left(1+\delta_0 / R e_p^{1 / 2}\right)^2$) with the measured drag coefficient ($C_{De,\mathrm{mean}}$), as discussed in Section \ref{sec:31}, it is necessary to calculate the maximum projected area, $A_{e,\mathrm{max}}$. Given that our snow particle analyzer only measures the general projected area ($A_e$) and the circumscribed area ($A$) at an unknown orientation, we assume that the ratio $A/A_e$ remains constant regardless of orientation. Consequently, the maximum projected area can be estimated as $A_{e,\mathrm{max}} = A_\mathrm{max}(A_e / A)$. In this equation, the maximum circumscribed area is calculated as $A_\mathrm{max} = \pi D_\mathrm{maj}^2 / 4$ for plates and dendrites and as $A_\mathrm{max} = \pi D_\mathrm{maj} D_\mathrm{min} / 4$ for other snow particle types, where $D_\mathrm{maj}$ and $D_\mathrm{min}$ are measured by the snow particle analyzer.

\bibliography{JFM_snow_settling_arxiv}

\end{document}